\pdfoutput=1

\documentclass[12pt,a4paper]{article}

\usepackage{ifthen}
\newboolean{pdflatex}
\setboolean{pdflatex}{true}

\newboolean{articletitles}
\setboolean{articletitles}{true}

\newboolean{uprightparticles}
\setboolean{uprightparticles}{false}

\newboolean{inbibliography}
\setboolean{inbibliography}{false}

\usepackage{microtype}
\usepackage{lineno}
\usepackage{xspace}

\usepackage{caption}

\usepackage{graphicx}
\usepackage{color}
\usepackage{colortbl}
\graphicspath{{./figs/}}

\usepackage{amsmath}
\usepackage{amssymb}
\usepackage{amsfonts}
\usepackage{upgreek}

\newcommand*\patchAmsMathEnvironmentForLineno[1]{%
\expandafter\let\csname old#1\expandafter\endcsname\csname #1\endcsname
\expandafter\let\csname oldend#1\expandafter\endcsname\csname
end#1\endcsname
 \renewenvironment{#1}%
   {\linenomath\csname old#1\endcsname}%
   {\csname oldend#1\endcsname\endlinenomath}%
}
\newcommand*\patchBothAmsMathEnvironmentsForLineno[1]{%
  \patchAmsMathEnvironmentForLineno{#1}%
  \patchAmsMathEnvironmentForLineno{#1*}%
}
\AtBeginDocument{%
\patchBothAmsMathEnvironmentsForLineno{equation}%
\patchBothAmsMathEnvironmentsForLineno{align}%
\patchBothAmsMathEnvironmentsForLineno{flalign}%
\patchBothAmsMathEnvironmentsForLineno{alignat}%
\patchBothAmsMathEnvironmentsForLineno{gather}%
\patchBothAmsMathEnvironmentsForLineno{multline}%
}

\usepackage{hyperref}
\usepackage[all]{hypcap}

\def\lhcb {\mbox{LHCb}\xspace}

\def\cdf    {\mbox{CDF}\xspace}
\def\dzero  {\mbox{D0}\xspace}
\def\aleph  {\mbox{ALEPH}\xspace}
\def\delphi {\mbox{DELPHI}\xspace}

\ifthenelse{\boolean{uprightparticles}}%
{

 \def\Pmu         {\ensuremath{\upmu}\xspace}

 \def\Ppi         {\ensuremath{\uppi}\xspace}

 \def\Ppsi        {\ensuremath{\uppsi}\xspace}

 \def\PDelta      {\ensuremath{\Delta}\xspace}                 
 \def\PXi      {\ensuremath{\Xi}\xspace}                 
 \def\PLambda      {\ensuremath{\Lambda}\xspace}                 
 \def\PSigma      {\ensuremath{\Sigma}\xspace}                 
 \def\POmega      {\ensuremath{\Omega}\xspace}                 
 \def\PUpsilon      {\ensuremath{\Upsilon}\xspace}                 
 
 \def\PB      {\ensuremath{\mathrm{B}}\xspace}                 
                  
 \def\PD      {\ensuremath{\mathrm{D}}\xspace}

 \def\PJ      {\ensuremath{\mathrm{J}}\xspace}                 
 \def\PK      {\ensuremath{\mathrm{K}}\xspace}

 \def\Pb      {\ensuremath{\mathrm{b}}\xspace}                 
 \def\Pc      {\ensuremath{\mathrm{c}}\xspace}

 \def\Pi      {\ensuremath{\mathrm{i}}\xspace}

}
{

 \def\Pmu         {\ensuremath{\mu}\xspace}

 \def\Ppi         {\ensuremath{\pi}\xspace}

 \def\Ppsi        {\ensuremath{\psi}\xspace}                 
                  
 \mathchardef\PDelta="7101
 \mathchardef\PXi="7104
 \mathchardef\PLambda="7103
 \mathchardef\PSigma="7106
 \mathchardef\POmega="710A
 \mathchardef\PUpsilon="7107
                  
 \def\PB      {\ensuremath{B}\xspace}                 
                  
 \def\PD      {\ensuremath{D}\xspace}

 \def\PJ      {\ensuremath{J}\xspace}                 
 \def\PK      {\ensuremath{K}\xspace}

 \def\Pb      {\ensuremath{b}\xspace}                 
 \def\Pc      {\ensuremath{c}\xspace}

 \def\Pi      {\ensuremath{i}\xspace}

}

\def\mup        {{\ensuremath{\Pmu^+}}\xspace}
\def\mun        {{\ensuremath{\Pmu^-}}\xspace} 

\def\cquark    {{\ensuremath{\Pc}}\xspace}

\def\bquark    {{\ensuremath{\Pb}}\xspace}

\def\pion   {{\ensuremath{\Ppi}}\xspace}

\def\pim    {{\ensuremath{\pion^-}}\xspace}

\def\kaon    {{\ensuremath{\PK}}\xspace}
  \def\Kbar    {{\kern 0.2em\overline{\kern -0.2em \PK}{}}\xspace}

\def\Km      {{\ensuremath{\kaon^-}}\xspace}

\def\KS      {{\ensuremath{\kaon^0_{\rm\scriptscriptstyle S}}}\xspace}

  \def\Dbar    {{\kern 0.2em\overline{\kern -0.2em \PD}{}}\xspace}

\def\B       {{\ensuremath{\PB}}\xspace}
\def\Bbar    {{\ensuremath{\kern 0.18em\overline{\kern -0.18em \PB}{}}}\xspace}

\def\Bz      {{\ensuremath{\B^0}}\xspace}

\def\jpsi     {{\ensuremath{{\PJ\mskip -3mu/\mskip -2mu\Ppsi\mskip 2mu}}}\xspace}

  \def\Y#1S{\ensuremath{\PUpsilon{(#1S)}}\xspace}

\def\Lbar        {{\ensuremath{\kern 0.1em\overline{\kern -0.1em\PLambda}}}\xspace}

\def\myL {\ensuremath{\PLambda}\xspace}
\def\myLb {\ensuremath{\PLambda^0_\bquark}\xspace}
\def\myX {\ensuremath{\PXi^-}\xspace}

\def\myXb {\ensuremath{\PXi^-_\bquark}\xspace}
\def\myO {\ensuremath{\POmega^-}\xspace}
\def\myOb {\ensuremath{\POmega^-_\bquark}\xspace}
\def\myXbz {\ensuremath{\PXi^0_\bquark}\xspace}

\def\to                 {\ensuremath{\rightarrow}\xspace}

\def\AT#1     {\ensuremath{A_{\mathrm{T}}^{#1}}\xspace}           

\def\C#1      {\ensuremath{\mathcal{C}_{#1}}\xspace}                       
\def\Cp#1     {\ensuremath{\mathcal{C}_{#1}^{'}}\xspace}                    
\def\Ceff#1   {\ensuremath{\mathcal{C}_{#1}^{\mathrm{(eff)}}}\xspace}        
\def\Cpeff#1  {\ensuremath{\mathcal{C}_{#1}^{'\mathrm{(eff)}}}\xspace}       
\def\Ope#1    {\ensuremath{\mathcal{O}_{#1}}\xspace}                       
\def\Opep#1   {\ensuremath{\mathcal{O}_{#1}^{'}}\xspace}                    

\newcommand{\tev}{\ifthenelse{\boolean{inbibliography}}{\ensuremath{~T\kern -0.05em eV}\xspace}{\ensuremath{\mathrm{\,Te\kern -0.1em V}}}\xspace}
\newcommand{\gev}{\ensuremath{\mathrm{\,Ge\kern -0.1em V}}\xspace}
\newcommand{\mev}{\ensuremath{\mathrm{\,Me\kern -0.1em V}}\xspace}
\newcommand{\kev}{\ensuremath{\mathrm{\,ke\kern -0.1em V}}\xspace}
\newcommand{\ev}{\ensuremath{\mathrm{\,e\kern -0.1em V}}\xspace}
\newcommand{\gevc}{\ensuremath{{\mathrm{\,Ge\kern -0.1em V\!/}c}}\xspace}
\newcommand{\mevc}{\ensuremath{{\mathrm{\,Me\kern -0.1em V\!/}c}}\xspace}
\newcommand{\gevcc}{\ensuremath{{\mathrm{\,Ge\kern -0.1em V\!/}c^2}}\xspace}
\newcommand{\gevgevcccc}{\ensuremath{{\mathrm{\,Ge\kern -0.1em V^2\!/}c^4}}\xspace}
\newcommand{\mevcc}{\ensuremath{{\mathrm{\,Me\kern -0.1em V\!/}c^2}}\xspace}

\def\mum  {\ensuremath{{\,\upmu\rm m}}\xspace}

\def\invfb   {\ensuremath{\mbox{\,fb}^{-1}}\xspace}

\def\gsim{{~\raise.15em\hbox{$>$}\kern-.85em
          \lower.35em\hbox{$\sim$}~}\xspace}
\def\lsim{{~\raise.15em\hbox{$<$}\kern-.85em
          \lower.35em\hbox{$\sim$}~}\xspace}

\def\ptot       {\mbox{$p$}\xspace}
\def\pt         {\mbox{$p_{\rm T}$}\xspace}

\def\evtgen     {\mbox{\textsc{EvtGen}}\xspace}

\def\geant      {\mbox{\textsc{Geant4}}\xspace}

\def\photos     {\mbox{\textsc{Photos}}\xspace}

\def\pythia     {\mbox{\textsc{Pythia}}\xspace}

\def\tell1  {TELL1\xspace}
\def\ukl1   {UKL1\xspace}

\usepackage{cite}
\usepackage{mciteplus}

\begin{document}

\renewcommand{\thefootnote}{\fnsymbol{footnote}}
\setcounter{footnote}{1}

\begin{titlepage}
\pagenumbering{roman}

\vspace*{-1.5cm}
\centerline{\large EUROPEAN ORGANIZATION FOR NUCLEAR RESEARCH (CERN)}
\vspace*{1.5cm}
\hspace*{-0.5cm}
\begin{tabular*}{\linewidth}{lc@{\extracolsep{\fill}}r}
\ifthenelse{\boolean{pdflatex}}
{\vspace*{-2.7cm}\mbox{\!\!\!\includegraphics[width=.14\textwidth]{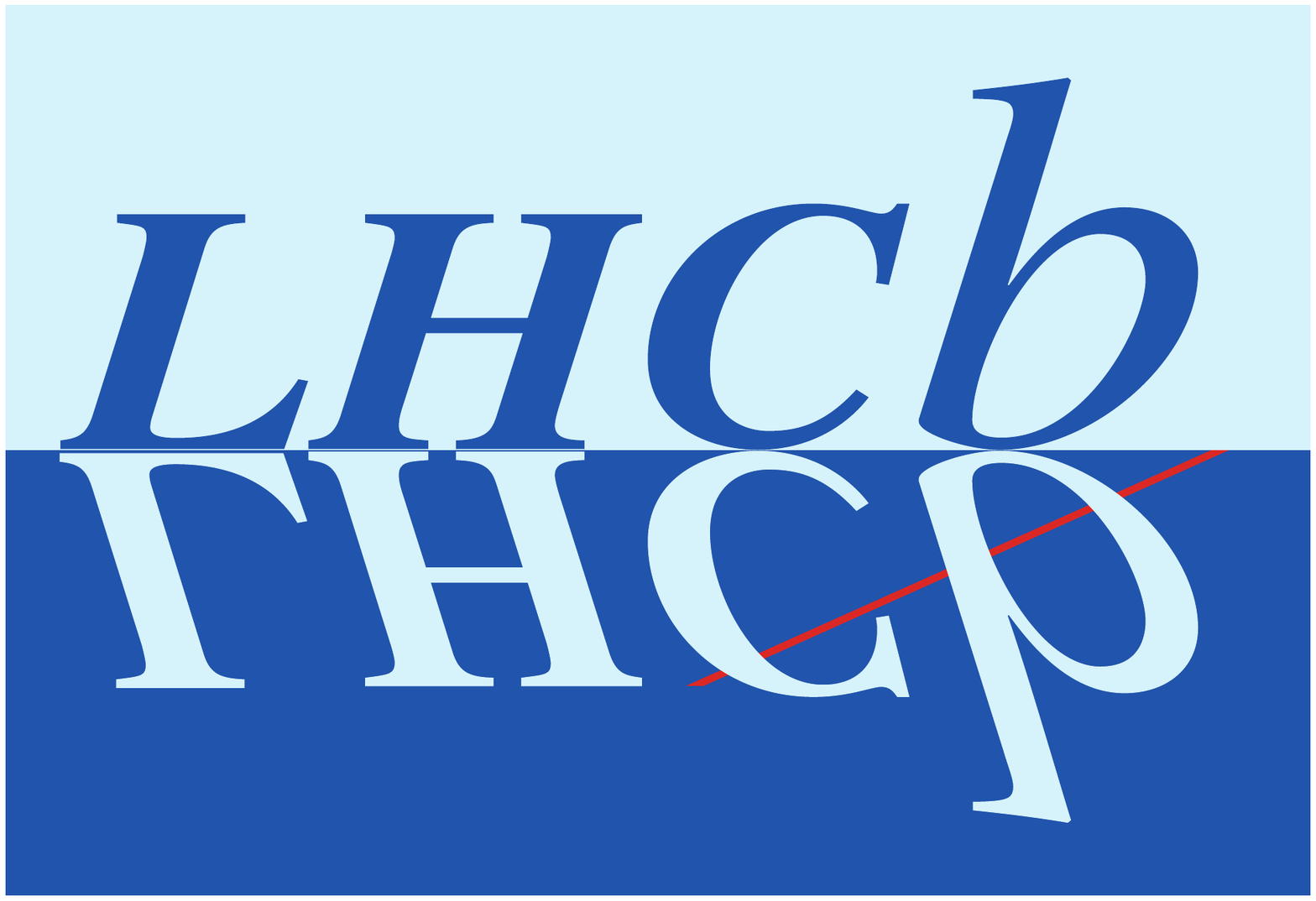}} & &}%
{\vspace*{-1.2cm}\mbox{\!\!\!\includegraphics[width=.12\textwidth]{lhcb-logo.eps}} & &}%
\\
 & & CERN-PH-EP-2014-081 \\
 & & LHCb-PAPER-2014-010 \\
 & & May 7, 2014 / revised June 18, 2014 \\
 & & \\
\end{tabular*}

\vspace*{4.0cm}

{\bf\boldmath\huge
\begin{center}
  Measurement of the \myXb and \myOb baryon lifetimes
\end{center}
}

\vspace*{2.0cm}

\begin{center}
The LHCb collaboration\footnote{Authors are listed on the following pages.}
\end{center}

\vspace{\fill}

\begin{abstract}
  \noindent
Using a data sample of $pp$ collisions corresponding to an integrated luminosity of 3\,fb$^{-1}$, the \myXb and \myOb baryons are reconstructed in the $\myXb \to \jpsi \myX$ and $\myOb \to \jpsi \myO$ decay modes and their lifetimes measured to be
\begin{eqnarray*}
\tau (\myXb) & = & 1.55\, ^{+0.10}_{-0.09}~{\rm(stat)} \pm 0.03\,{\rm(syst)\,ps}, \\
\tau (\myOb) & = & 1.54\, ^{+0.26}_{-0.21}~{\rm(stat)} \pm 0.05\,{\rm(syst)\,ps}.
\end{eqnarray*}
These are the most precise determinations to date. Both measurements are in good agreement with previous experimental results and with theoretical predictions.
\end{abstract}

\vspace*{2.0cm}

\begin{center}
Published in Phys. Lett. B 736, 154-162 (2014)
\end{center}

\vspace{\fill}

{\footnotesize 
\centerline{\copyright~CERN on behalf of the \lhcb collaboration, license \href{http://creativecommons.org/licenses/by/3.0/}{CC-BY-3.0}.}}
\vspace*{2mm}

\end{titlepage}

\newpage
\setcounter{page}{2}
\mbox{~}
\newpage

\centerline{\large\bf LHCb collaboration}
\begin{flushleft}
\small
R.~Aaij$^{41}$, 
B.~Adeva$^{37}$, 
M.~Adinolfi$^{46}$, 
A.~Affolder$^{52}$, 
Z.~Ajaltouni$^{5}$, 
J.~Albrecht$^{9}$, 
F.~Alessio$^{38}$, 
M.~Alexander$^{51}$, 
S.~Ali$^{41}$, 
G.~Alkhazov$^{30}$
% ...

{\footnotesize \it
% ...
$ ^{5}$Clermont Universit\'{e}, Universit\'{e} Blaise Pascal, CNRS/IN2P3, LPC, Clermont-Ferrand, France\\
% ...
$ ^{9}$Fakult\"{a}t Physik, Technische Universit\"{a}t Dortmund, Dortmund, Germany\\
% ...
$ ^{30}$Petersburg Nuclear Physics Institute (PNPI), Gatchina, Russia\\
% ...
$ ^{37}$Universidad de Santiago de Compostela, Santiago de Compostela, Spain\\
$ ^{38}$European Organization for Nuclear Research (CERN), Geneva, Switzerland\\
% ...
$ ^{41}$Nikhef National Institute for Subatomic Physics, Amsterdam, The Netherlands\\
% ...
$ ^{46}$H.H. Wills Physics Laboratory, University of Bristol, Bristol, United Kingdom\\
% ...
$ ^{51}$School of Physics and Astronomy, University of Glasgow, Glasgow, United Kingdom\\
$ ^{52}$Oliver Lodge Laboratory, University of Liverpool, Liverpool, United Kingdom\\
% ...
}
\end{flushleft}

\cleardoublepage

\renewcommand{\thefootnote}{\arabic{footnote}}
\setcounter{footnote}{0}

\pagestyle{plain}
\setcounter{page}{1}
\pagenumbering{arabic}


\section{Introduction}

Heavy baryons are systems of three quarks, among which at least one is $c$ or $b$. The quarks are bound by the strong interaction, which is described by quantum chromodynamics (QCD). Hadron lifetimes are among the most useful inputs to tune the parameters of QCD models. A powerful approach for theoretical predictions of $b$-hadron lifetime ratios is the heavy quark expansion (HQE) framework~\cite{Neubert:1998qx} which allows calculations in powers of $\Lambda_{\rm QCD}/m_b$, where $\Lambda_{\rm QCD}$ is the energy scale at which QCD becomes non-perturbative and $m_b$ is the $b$-quark mass. With the exception of the $b$ hadrons containing a $c$ quark, the predictions for the various $b$-hadron lifetimes only start to differ at the order $\Lambda_{\rm QCD}^2/m_b^2$ and are equal within several percent.

So far only the most abundantly produced $b$ baryon, the \myLb with quark content $udb$, has been studied in detail. Early \myLb lifetime measurements~\cite{Buskulic:1995px, Akers:1995wl, Abreu:1996py, Abe:1996xl} yielded values significantly smaller than the $B$-meson lifetime determinations, casting doubt on the HQE and causing increased theoretical activity~\cite{Altarelli:1996zb, Cheng:1997, Ito:1998, Uraltsev:2000vt, Gabbiani:2003xx, Gabbiani:2004ya}. More recent determinations of the ratio between the \myLb and \Bz lifetimes, for instance that of Ref.~\cite{LHCb-PAPER-2014-003}, are in much better agreement with the original predictions. However, less information exists on the strange $b$ baryons, which are less abundantly produced than \myLb baryons. In particular for the \myXb ($dsb$) and \myOb ($ssb$) baryons only a few theoretical lifetime calculations are available~\cite{Bigi:1995, Cheng:1997, Ito:1998}. Furthermore, most of the predictions date back to the 1990s and have limited precision, with central values ranging from 1.0\,ps to 1.7\,ps. New experimental data are needed to provide more stringent constraints on the models.

The weakly decaying \myXb and \myOb baryons were observed for the first time at the Tevatron experiments \cdf~\cite{Aaltonen:2007ap, Aaltonen:2009ny} and \dzero~\cite{Abazov:2007am, Abazov:2008qm}. Prior to these first observations, the average $\PXi_{b}$ lifetime (including \myXb and \myXbz) was measured by the LEP experiments \delphi~\cite{Abreu:1995qq, Abdallah:2005xi} and \aleph~\cite{Buskulic:1996zy} using partially reconstructed decays. So far the only exclusive lifetime measurement of the strange $b$ baryons \myXb and \myOb has been made by the \cdf experiment~\cite{Aaltonen:2009ny, Aaltonen:2014al}. Recently, \lhcb demonstrated its ability to reconstruct a significant number of \myXb and \myOb baryons~\cite{LHCb-PAPER-2012-048} and to measure precisely $b$-hadron lifetimes~\cite{LHCb-PAPER-2013-065}.

In this Letter we present lifetime measurements of the \myXb and \myOb baryons reconstructed in the $\myXb \to \jpsi \myX$ and $\myOb \to \jpsi \myO$ decay modes. The daughter particles are reconstructed in the decay modes $\jpsi \to \mup \mun$, $\myX \to \myL \pim$, $\myO \to \myL \Km$ and $\myL \to p \pim$. Unless specified otherwise, charge-conjugated states are implied throughout.

\section{Detector and event samples}
\label{sec:Detector}

The \lhcb detector~\cite{Alves:2008zz} is a single-arm forward
spectrometer covering the \mbox{pseudorapidity} range $2<\eta <5$,
designed for the study of particles containing \bquark or \cquark
quarks. The detector includes a high-precision tracking system
consisting of a silicon-strip vertex detector surrounding the $pp$
interaction region, a large-area silicon-strip detector located
upstream of a dipole magnet with a bending power of about
$4{\rm\,Tm}$, and three stations of silicon-strip detectors and straw
drift tubes~\cite{LHCb-DP-2013-003} placed downstream of the magnet.
The combined tracking system provides a momentum measurement with
a relative uncertainty that varies from 0.4\% at low momentum, \ptot, to 0.6\% at 100\gevc,
and an impact parameter measurement with a resolution of 20\mum for
charged particles with large transverse momentum, \pt. Different types of charged hadrons are distinguished using information
from two ring-imaging Cherenkov detectors~\cite{LHCb-DP-2012-003}. Photon, electron and
hadron candidates are identified by a calorimeter system consisting of
scintillating-pad and preshower detectors, an electromagnetic
calorimeter and a hadronic calorimeter. Muons are identified by a
system composed of alternating layers of iron and multiwire
proportional chambers~\cite{LHCb-DP-2012-002}.

The trigger consists of a
hardware stage, based on information from the calorimeter and muon
systems, followed by a software stage, which applies a full event
reconstruction. For this measurement, events are first required to pass the hardware trigger,
which selects muons with high transverse momentum.
In the subsequent software stage, events are retained by two independent sets of requirements.
One demands a muon candidate with momentum larger than 6\gevc that,
combined with another oppositely charged muon candidate, yields a dimuon mass larger than 2.7\gevcc.
The other requires a muon candidate with momentum larger than 8\gevc
and an impact parameter above $100\mum$ with respect to all
of the primary $pp$ interaction vertices~(PVs) in the
event. Finally, for all candidates, two 
muons are required to form a vertex that is significantly
displaced from the PVs.

The \myXb and \myOb lifetime measurements presented here are based on the combination of the two data sets recorded in 2011 and 2012. During the year 2011 the \lhcb detector recorded $pp$ collisions at a centre-of-mass energy of $\sqrt{s} = 7\tev$ corresponding to an integrated luminosity of 1\,fb$^{-1}$. In 2012, it recorded approximately twice as much data at $\sqrt{s} = 8\tev$. Between 2011 and 2012, an improvement in the tracking algorithm of the vertex detector was introduced, leading to different trigger and reconstruction efficiencies in the two data sets. The polarity of the dipole magnet was periodically inverted so that roughly half of the data was collected with each polarity.

Four million \myXb (\myOb) signal events, corresponding to approximately 135\,\invfb (1700\,\invfb) of \lhcb data, were simulated with each of the 2011 and 2012 data taking conditions. The $pp$ collisions are generated using \pythia~\cite{Sjostrand:2006za,*Sjostrand:2007gs} with a specific \lhcb configuration~\cite{LHCb-PROC-2010-056}. Decays of hadronic particles are described by \evtgen~\cite{Lange:2001uf}, in which final state radiation is generated using \photos~\cite{Golonka:2005pn}. The interaction of the generated particles with the detector and its response are implemented using the \geant toolkit~\cite{Allison:2006ve, *Agostinelli:2002hh} as described in Ref.~\cite{LHCb-PROC-2011-006}.

\section{Reconstruction and selection}
\label{Sec:Selection}

The $\jpsi \to \mup \mun$ decay is reconstructed from oppositely charged particles that leave deposits in the vertex detector, the tracking stations and the muon system. The hyperons in the $b$-baryon decay chains (\myX, \myO and \myL) are long-lived; approximately 10$\%$ of all reconstructed $b$-baryon candidates are reconstructed with all tracks leaving deposits in the vertex detector. To retain as many candidates as possible, tracks that have no vertex detector information are also considered for the reconstruction of the hyperon decays.

The \myXb and \myOb candidates are selected through identical requirements except for the ranges in which the baryon masses are reconstructed. In addition, for the \myOb case the charged track from the \myO decay is required to be identified as a kaon by the particle identification detectors, removing more than 95\% of the background pions.

All final-state tracks are required to satisfy minimal quality criteria and kinematic requirements. In order to reduce backgrounds from combinations of random tracks, the decay vertices are required to be well reconstructed. The $\jpsi$, $\myX$, $\myO$ and $\myL$ candidates are selected within mass windows of $\pm 60 \mevcc$, $\pm 11 \mevcc$, $\pm 11 \mevcc$ and $\pm 6 \mevcc$, respectively, around the corresponding known masses~\cite{PDG2012}.

The hadronic final-state tracks are required to have large impact parameters with respect to the PV associated with the $b$-baryon candidate. The associated PV is chosen as the PV giving the smallest increase in the $\chi^2$ of the PV fit when the $b$ baryon is included. The associated PV is also required to be isolated with respect to other PVs and consistent with the nominal interaction region.

The $b$-baryon mass is computed after a complete kinematic fit of the decay chain~\cite{Hulsbergen:2005pu} in which the masses of both daughter particles are constrained to their known values~\cite{PDG2012}. No constraint is applied on the \myL mass. The resulting $b$-baryon invariant mass is required to lie in the range 5600--6000\mevcc for \myXb candidates and 5800--6300\mevcc for \myOb candidates. The decay time of the $b$-baryon candidate, $t$, is computed from the decay length, $d$, as

\begin{equation}
t = \frac{d}{\beta \gamma c} = \frac{m}{p}\,d \, ,
\end{equation}

\noindent where $m$ is the reconstructed mass and $p$ the reconstructed momentum of the $b$-baryon candidate. The decay length itself is obtained from a refit of the decay chain with no mass constraints in order to keep the correlation between the reconstructed decay time and mass at a negligible level. Backgrounds are further suppressed by requiring this decay chain fit to be of good quality. The reconstructed decay time is required to lie in the range 0.3--14\,ps. The lower bound of this decay-time range helps to suppress background coming from random combinations of tracks with real \jpsi mesons produced at the PV. In less than 1\% of the cases, more than one candidate per event pass the selection criteria and only the candidate with the best decay chain fit result is retained.

\section{Resolution and efficiency}
\label{Sec:Acceptance}

The decay time resolution is obtained by fitting the difference between the reconstructed decay time, $t$, and the true decay time, $t_{\rm true}$, in simulated events. The fit model is a single Gaussian function $G(t-t_{\rm true}, \bar{t}, \sigma_{\rm res})$ where the mean, $\bar{t}$, and the width, $\sigma_{\rm res}$, are left free. For both considered decay modes and both data-taking periods, $\bar{t}$ is compatible with zero and $\sigma_{\rm res}$ is close to 50\,fs.

A bias in the measured lifetime can arise from a non-uniform efficiency as a function of the $b$-baryon decay time~\cite{LHCb-PAPER-2013-065}. There are two types of inefficiencies which alter the decay time distribution. The first affects mostly candidates with small decay times and is induced by the requirements of the trigger that reject predominantly short-lived $b$ baryons. The second affects mostly candidates with large decay times and is due to the geometrical detector acceptance, the reconstruction process and the selection criteria that lead to a lower efficiency for long-lived $b$ baryons. Both effects are estimated and corrected for using simulation. This approach is validated with several techniques described in Sec.~\ref{Sec:Systematics}.

\begin{figure}[h!]
\centering
\includegraphics[width=0.5\textwidth] {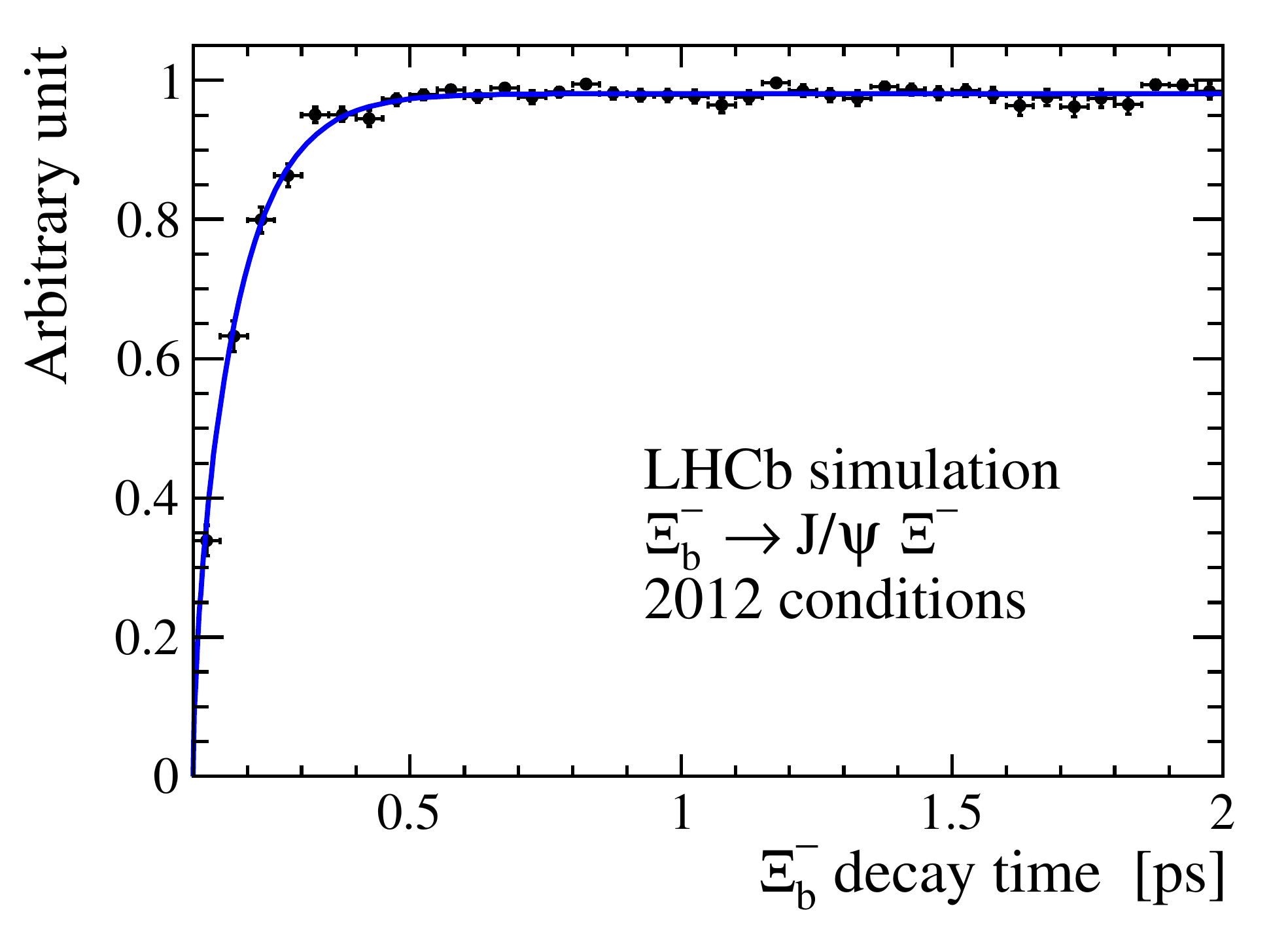}%
\includegraphics[width=0.5\textwidth] {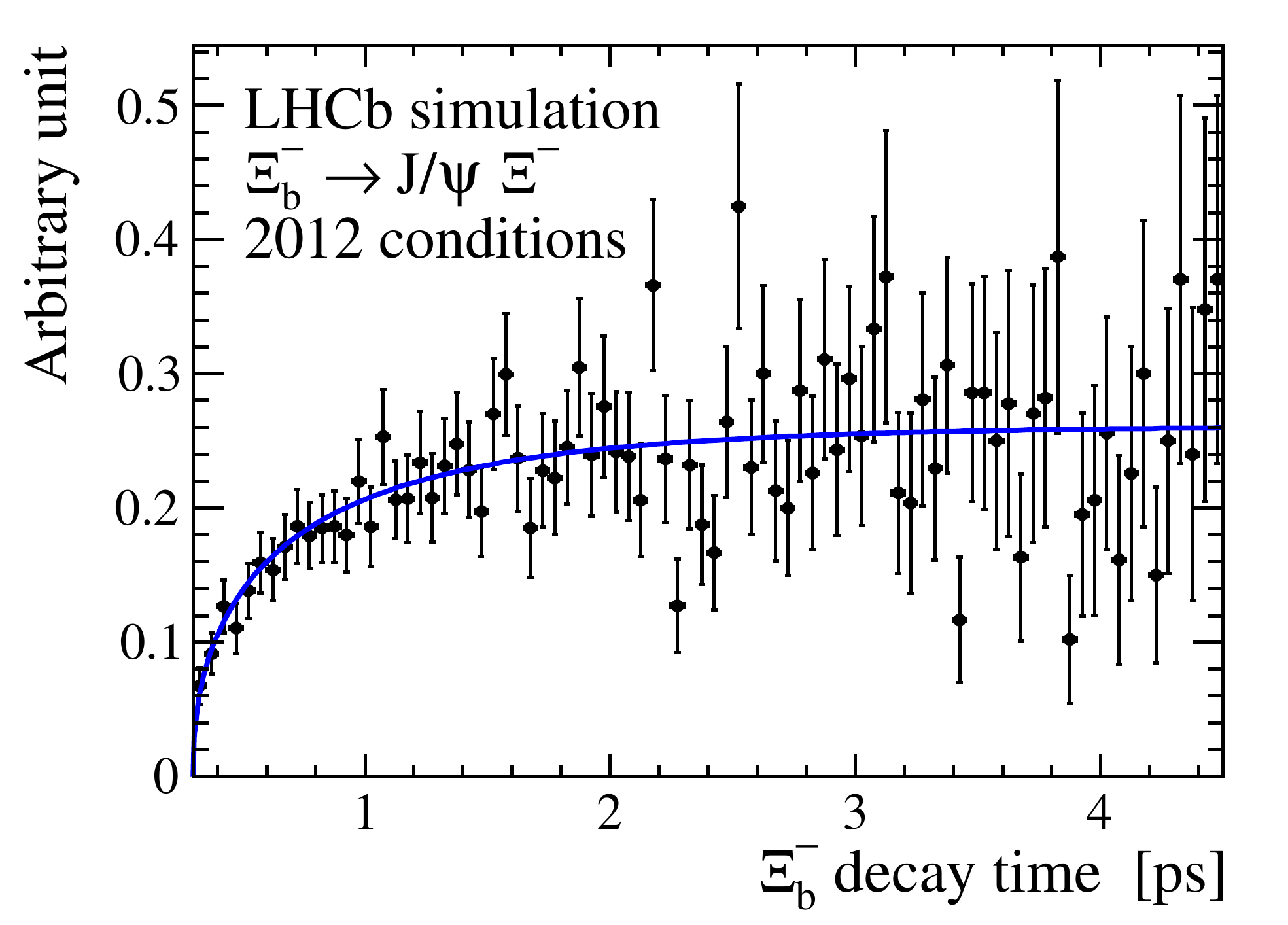} \\
\caption{\small Efficiency for triggering simulated \myXb events as a function of the reconstructed decay time under the 2012 data-taking conditions. Only a restricted decay-time range is shown in order to emphasize the region where the effect is large. The left panel shows the efficiency for events passing the trigger with only the requirement on the \jpsi vertex displacement. The right panel shows the efficiency for events passing the trigger with requirements on both the \jpsi vertex displacement and the muon impact parameter and required to not pass the other trigger. The results of fits with functions proportional to that given in Eq.~\ref{Eq:AcceptanceTrigger} are overlaid.}
\label{Fig:XibLowerAcceptance}
\end{figure}

The two trigger selections used for these lifetime measurements include a requirement on the decay length significance of the \jpsi meson. In addition, one selection also contains a requirement on the impact parameter of the muons from the \jpsi decay. The two requirements induce an inefficiency at low values of the reconstructed decay time. To assess this effect, simulated events undergo an emulation of the trigger. In addition, an unbiased trigger selection is used to remove the contribution from the detector acceptance, the reconstruction and the selection. The resulting efficiency as a function of the reconstructed decay time is fitted with an empirical function of the form

\begin{eqnarray}
\label{Eq:AcceptanceTrigger}
\varepsilon_1 (t) = {\rm erf}(\ a \cdot (t- t_0)^n \ ) \, , & {\rm where} & {\rm erf}(u) = \frac{2}{\sqrt{\pi}} \int_0^u e^{-x^2} dx \, ,
\end{eqnarray}

\noindent and where $a$, $t_0$ and $n$ are free parameters. The distributions of the decay products of the $b$ baryons depend on the decay mode and on the year of data taking. This dependence slightly affects the shape of the efficiency as a function of the reconstructed decay time. Thus separate efficiency functions are obtained for the two decay modes, for the two data taking periods and for the two trigger selections. The efficiency functions corresponding to the 2012 data taking conditions for the \myXb case are shown in Fig.~\ref{Fig:XibLowerAcceptance} as an example.

The dependence of the efficiency on the decay time due to the geometrical detector acceptance, the reconstruction and the selection is found to be well described with a linear function,

\begin{equation}
\label{Eq:AcceptanceReconstruction}
\varepsilon_2 (t) = 1 + \beta t \, .
\end{equation}

\begin{figure}[h!]
\centering
\includegraphics[width=0.5\textwidth] {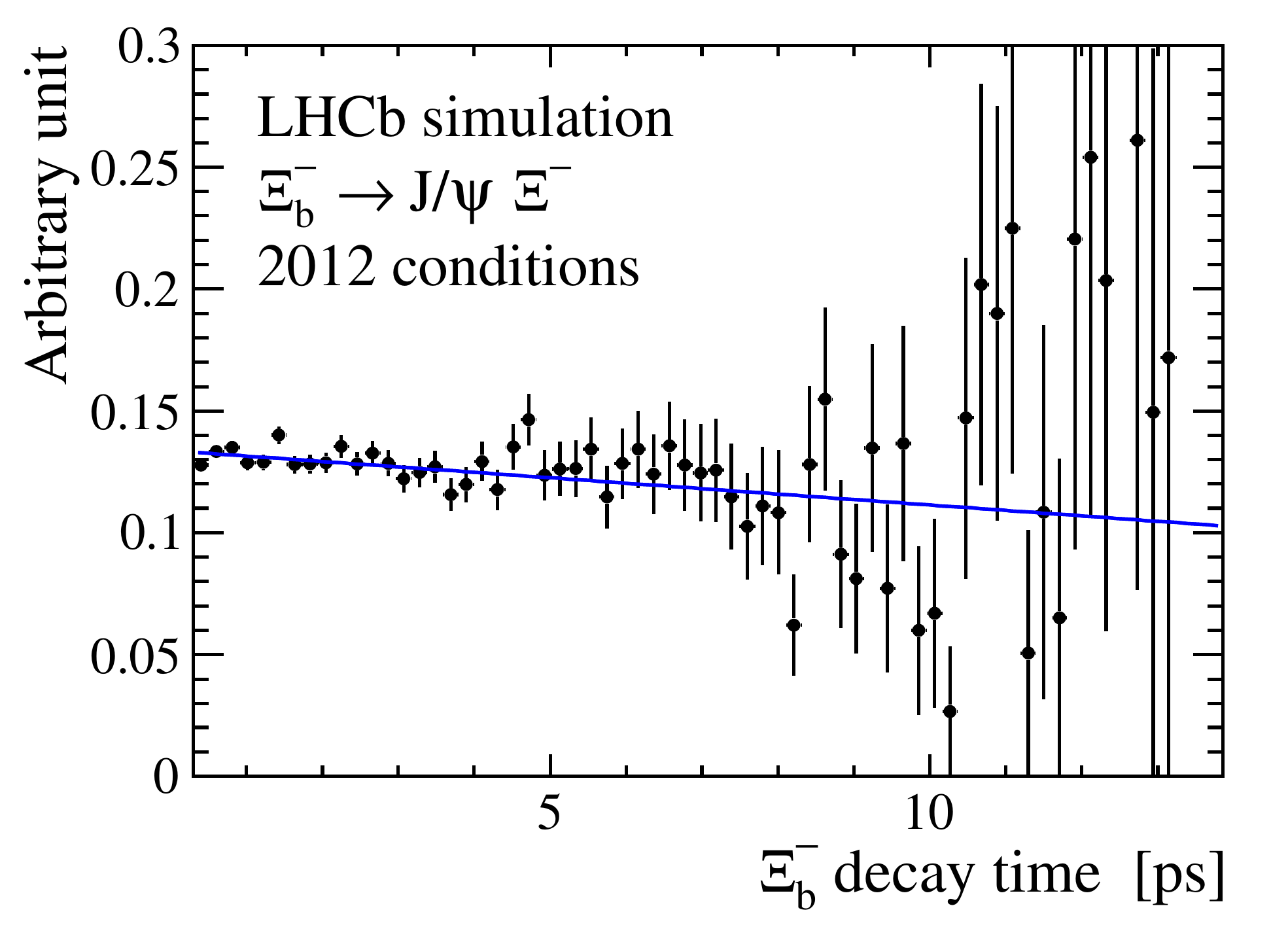}%
\includegraphics[width=0.5\textwidth] {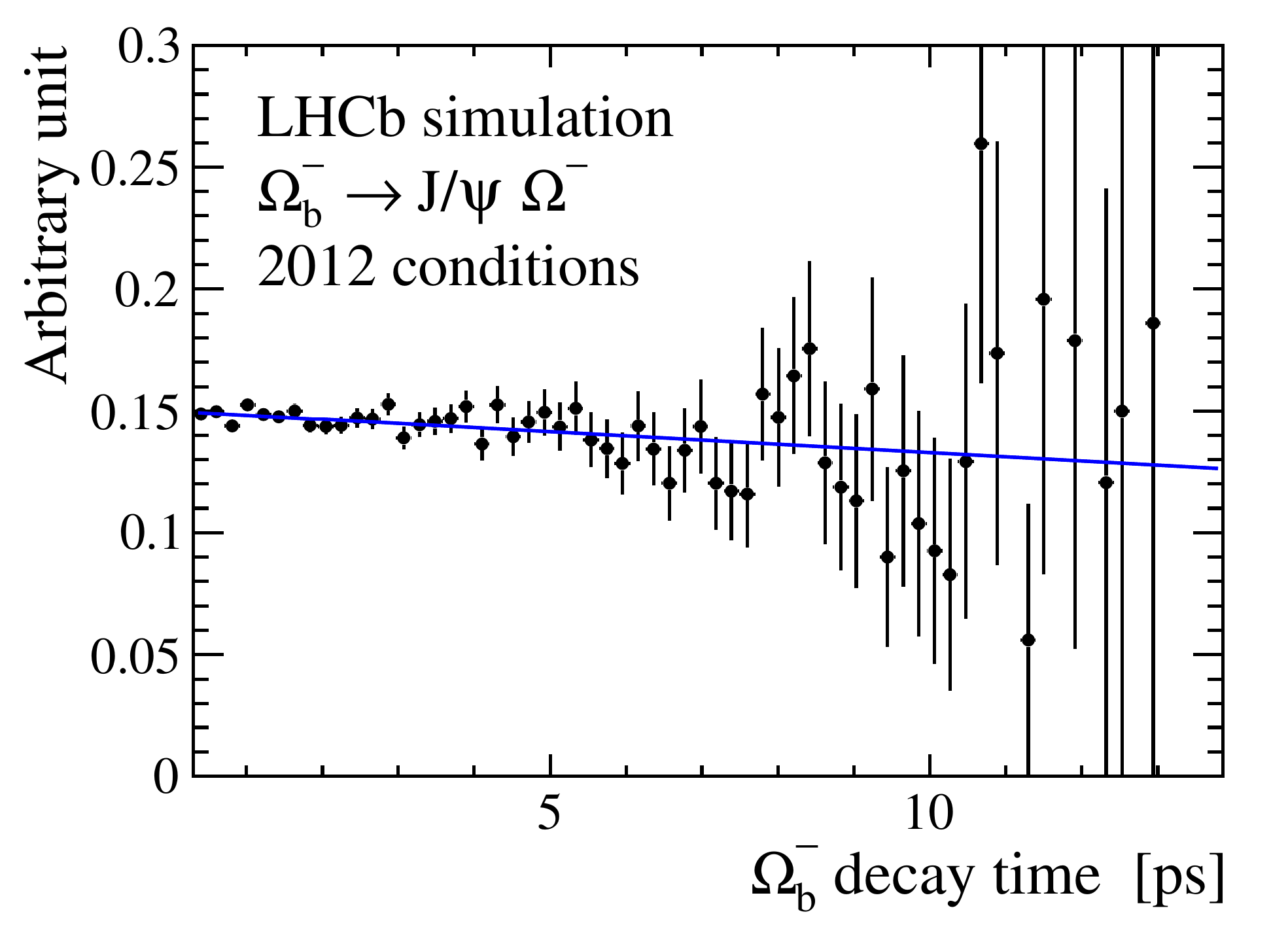} \\
\caption{\small Efficiency due to the detector acceptance, reconstruction and selection of simulated \myXb (left) and \myOb (right) events as a function of the reconstructed decay time under the 2012 data-taking conditions. The results of fits with functions proportional to that given in Eq.~\ref{Eq:AcceptanceReconstruction} are overlaid.}
\label{Fig:XibUpperAcceptance}
\end{figure}

\noindent The free parameter $\beta$ is obtained by fitting a function proportional to $\varepsilon_2 (t) \cdot \int_0^\infty \exp(-t_{\rm true}/\tau_{\rm gen}) \, G(t-t_{\rm true}, 0, \sigma_{\rm res}) \, dt_{\rm true}$ to the reconstructed decay time distribution of simulated signal events that are generated with a mean lifetime of $\tau_{\rm gen}$ and that are fully reconstructed and selected. Separate values for $\beta$ are determined for the two different decay modes and the two data-taking periods and are given in Table~\ref{Tab:BetaFactors}. The efficiency functions corresponding to the 2012 data taking conditions for the \myXb and \myOb cases are shown in Fig.~\ref{Fig:XibUpperAcceptance} as an example.

\clearpage

\begin{table}[h!]
\caption{\small Fitted $\beta$ values (in ps$^{-1}$) for the efficiency described in Eq.~\ref{Eq:AcceptanceReconstruction} extracted from simulated \myXb and \myOb decays. The quoted uncertainties are statistical only.}
\label{Tab:BetaFactors}
\begin{center}
\begin{tabular}{lcc}
\hline
Decay mode             & 2011 conditions                  & 2012 conditions                  \\
\hline
$\myXb \to \jpsi \myX$ & $(-13.1 \pm 4.8) \times 10^{-3}$ & $(-20.2 \pm 5.0) \times 10^{-3}$ \\
$\myOb \to \jpsi \myO$ & $(-23.3 \pm 3.5) \times 10^{-3}$ & $(-19.3 \pm 3.9) \times 10^{-3}$ \\
\hline
\end{tabular}
\end{center}
\end{table}

\section{Lifetime fit}
\label{Sec:Fit}
The lifetime is extracted from a two-dimensional extended maximum likelihood fit to the unbinned $b$-baryon mass and decay-time distributions. The mass and decay time are computed with the techniques described in Sec.~\ref{Sec:Selection}. Assuming a negligible correlation between these two quantities, the two-dimensional probability density functions for the signal and the background are each written as the product of a mass term and a decay-time term.

For the mass distribution, the signal is described with a single Gaussian function in which the mean and width are free parameters. Independent means are used for the data recorded in 2011 and in 2012, since different calibrations are applied. The background in the mass distribution is modelled with an exponential function. The signal in the decay time distribution is described with the product of the efficiency functions (described in Eqs.~\ref{Eq:AcceptanceTrigger}~and~\ref{Eq:AcceptanceReconstruction}) and a convolution between an exponential function and a Gaussian function describing the decay time resolution,

\begin{equation}
S(t) = N \cdot \varepsilon_1(t) \cdot \varepsilon_2(t) \cdot \int_0^\infty e^{-t_{\rm true}/\tau} G(t-t_{\rm true}, 0, \sigma_{\rm res}) \, dt_{\rm true} \, ,
\end{equation}

\noindent where $N$ is a normalisation parameter and $\tau$ the fitted lifetime. The decay time resolution $\sigma_{\rm res}$ is fixed to the value obtained in simulation, separately for each decay mode and each year of data taking. The background in the decay time distribution is modelled with the sum of two exponential functions that are also convolved with the fixed decay time resolution function. With the exception of $\sigma_{\rm res}$, all background parameters are left free in the fit. A study based on pseudo-experiments shows that no observable bias to the measured lifetimes arises from the fit model itself.

\begin{figure}[tb]
\centering
\includegraphics[width=0.5\textwidth] {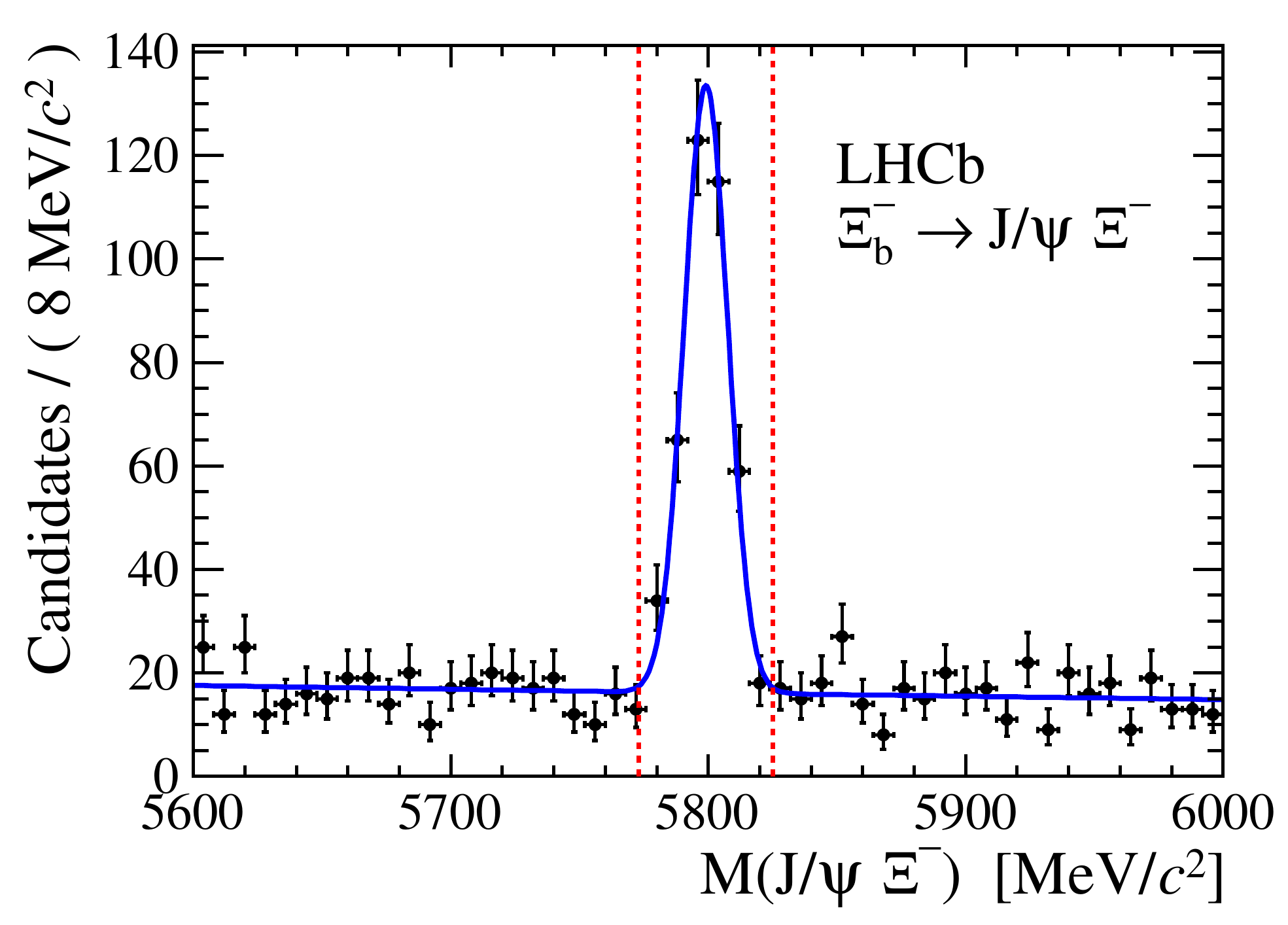}%
\includegraphics[width=0.5\textwidth] {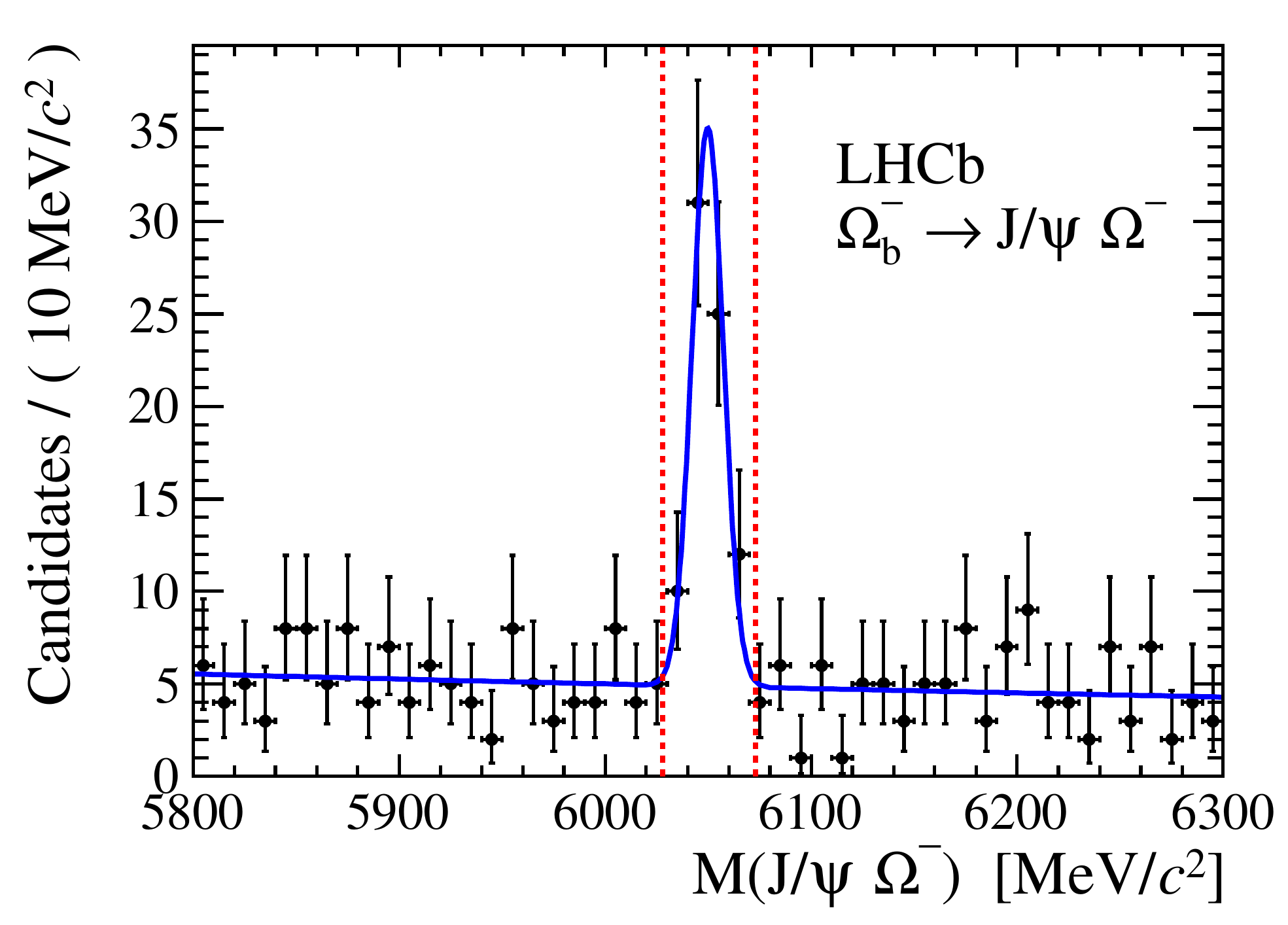}
\includegraphics[width=0.5\textwidth] {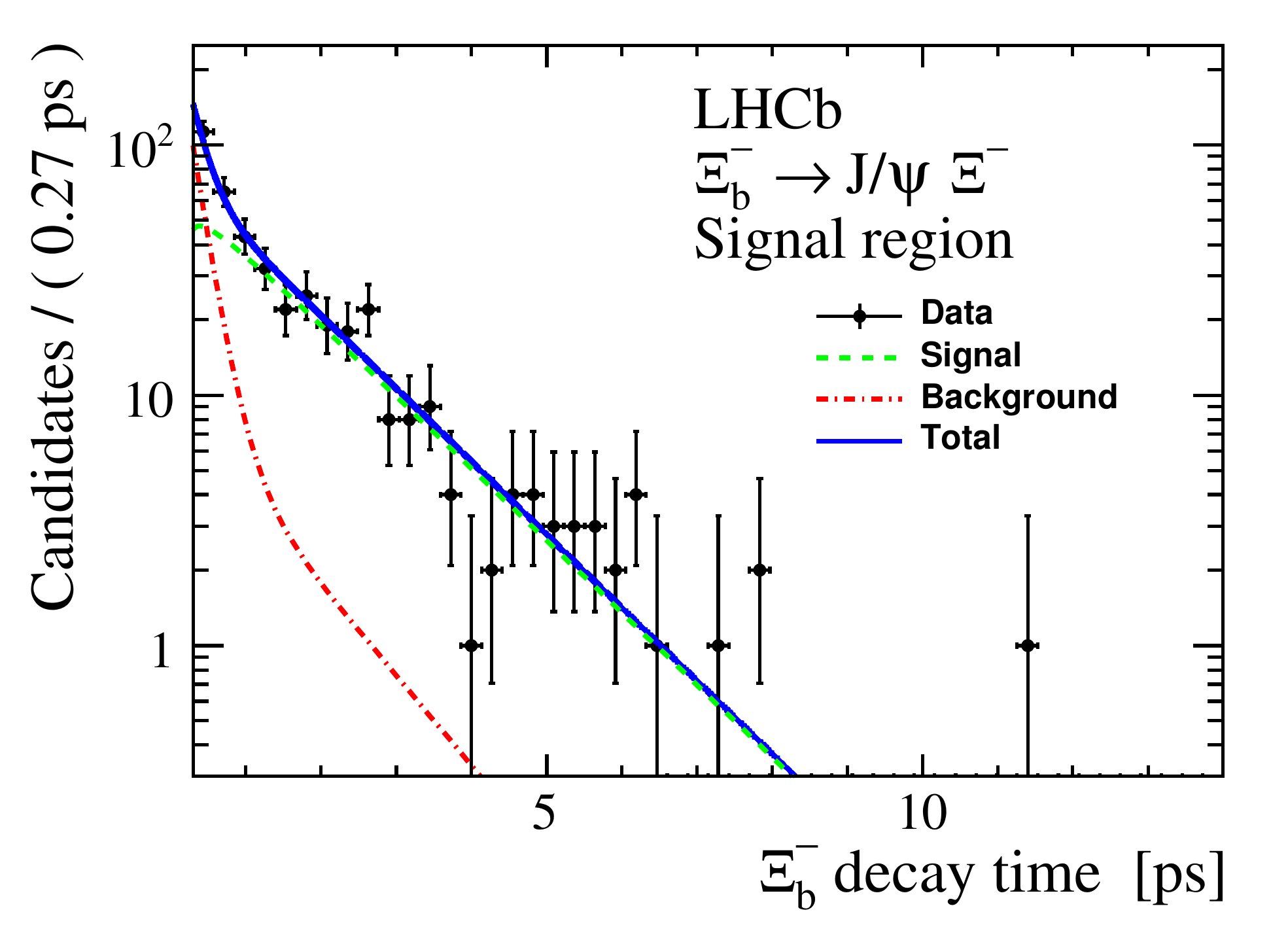}%
\includegraphics[width=0.5\textwidth] {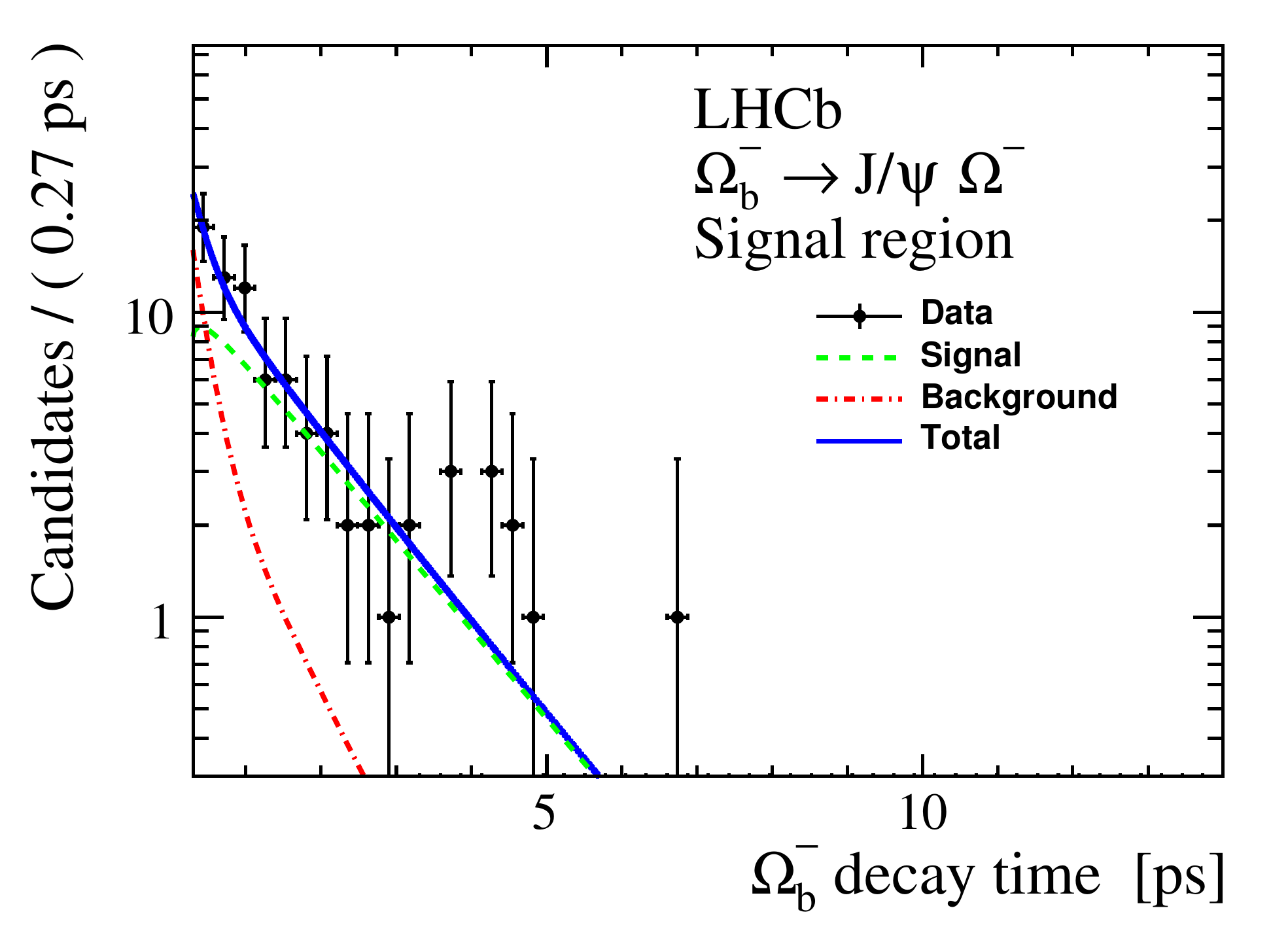}
\includegraphics[width=0.5\textwidth] {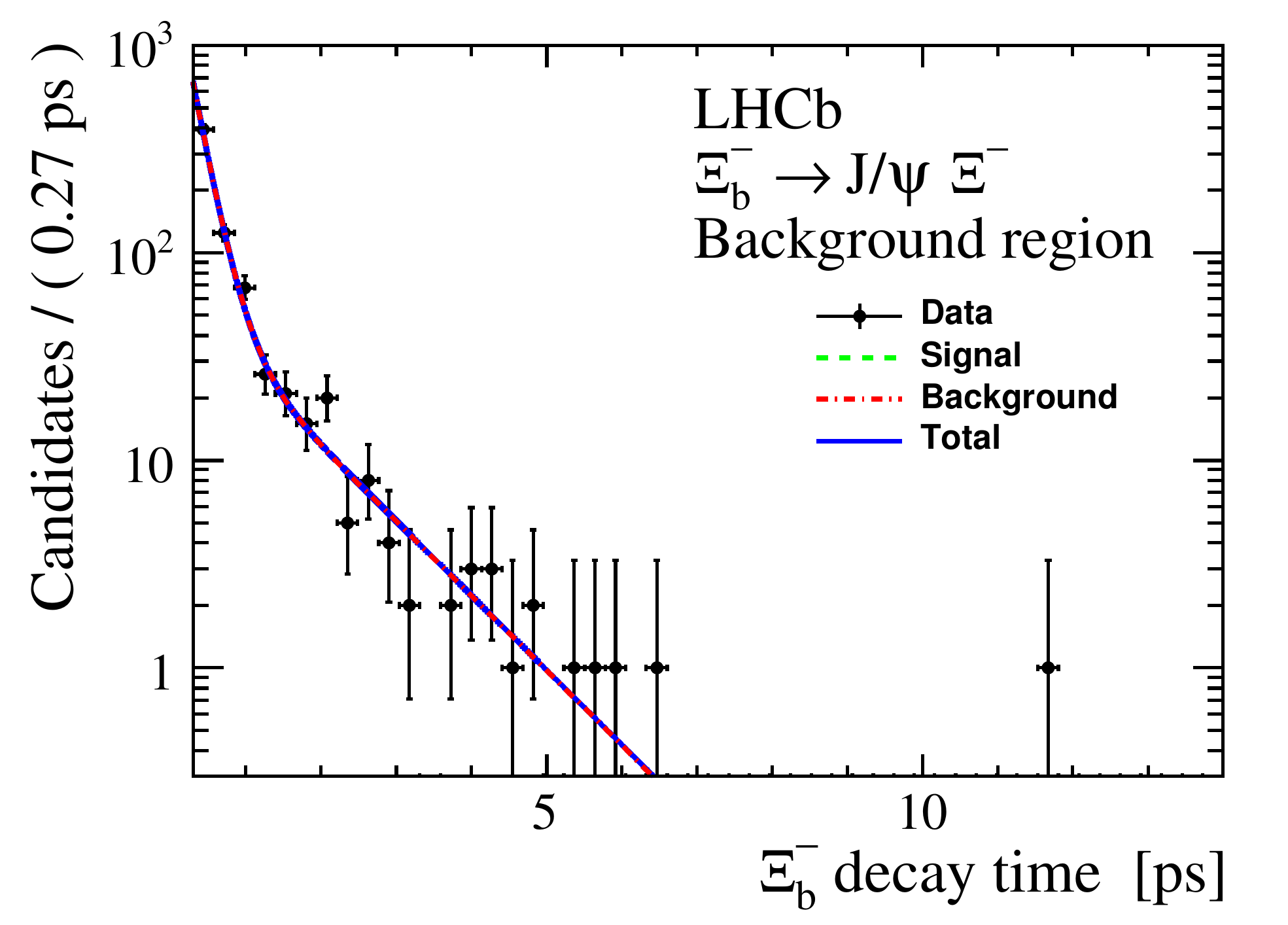}%
\includegraphics[width=0.5\textwidth] {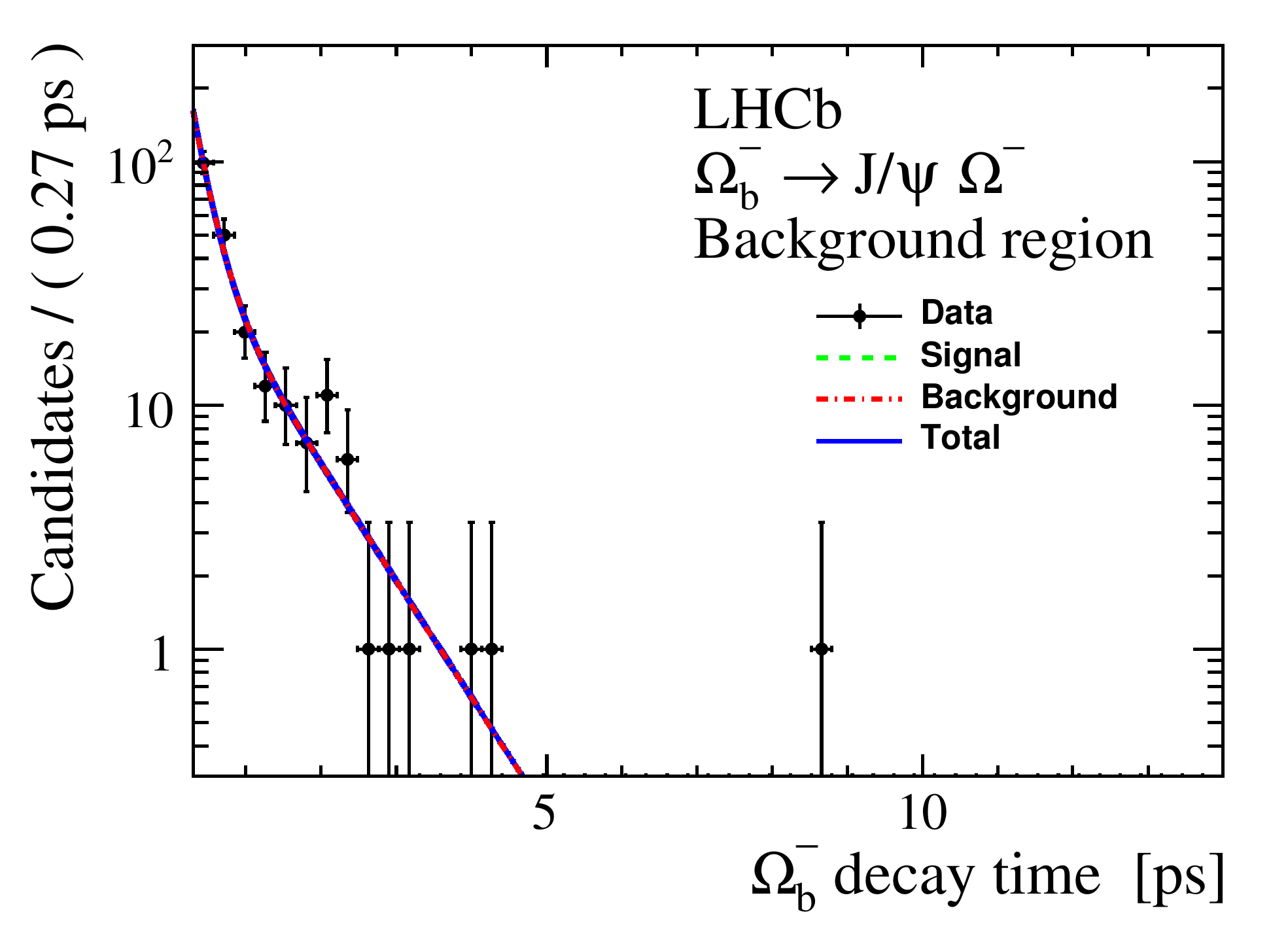}
\caption{\small Distributions of the reconstructed invariant mass (top) and decay time (middle and bottom) of the $\myXb\to\jpsi\myX$ (left) and $\myOb\to\jpsi\myO$ (right) candidates. The middle (bottom) panels show the decay time distributions of the candidates in the signal (background) mass regions. The signal mass region is defined as 5773--5825\mevcc for \myXb and 6028--6073\mevcc for \myOb candidates, as shown by the vertical dotted lines in the mass distributions, whereas the background mass regions include all other candidates. The results of the fits are overlaid.}
\label{Fig:LifetimeFit}
\end{figure}

The fit is performed for all selected $b$-baryon candidates. Due to the low signal yields, asymmetric uncertainties are calculated. Figure~\ref{Fig:LifetimeFit} shows the invariant mass and decay time distributions and the projection of the fit results for the \myXb and \myOb baryons. Table~\ref{Tab:Fit} displays the fit result for the relevant signal parameters.

As a consistency check for the fitting method, a measurement of the $\myLb \to \jpsi \myL$ lifetime is performed using the same data set and techniques as presented in this paper. The measured \myLb lifetime is consistent with the world average~\cite{PDG2012} and with recent measurements from LHCb~\cite{LHCb-PAPER-2013-065,LHCb-PAPER-2014-003}.

\clearpage

\begin{table}[t!]
\caption{\small Fitted parameters with statistical uncertainties for the \myXb and \myOb signals.}
\label{Tab:Fit}
\begin{center}
\begin{tabular}{lr@{\,}c@{\,}lr@{\,}c@{\,}l}
\hline
Parameter         & \multicolumn{3}{c}{$\myXb \to \jpsi \myX$} & \multicolumn{3}{c}{$\myOb \to \jpsi \myO$} \\
\hline
Signal yield      & 313  & $\pm$ & 20           & 58   & $\pm$ & 8            \\
Mass resolution   & 8.5  & $\pm$ & 0.5\,\mevcc  & 7.5  & $\pm$ & 1.0\,\mevcc  \\
Lifetime ($\tau$) & 1.55 & $^{+}_{-}$ & $^{0.10}_{0.09}$~ps & 1.54 & $^{+}_{-}$ & $^{0.26}_{0.21}$~ps \\
\hline
\end{tabular}
\end{center}
\end{table}

\section{Systematic uncertainties}
\label{Sec:Systematics}

Unless specified otherwise, the evaluation of the systematic uncertainties is performed by varying in turn each fixed parameter of the fit within its uncertainty and taking the change in the fit result. The total systematic uncertainty is obtained as the quadratic sum of the individual contributions. Distributions of results from fits to pseudo-experiments are used for the leading contributions (efficiencies and modelling) in order to ensure that they are not incorrectly estimated due to a statistical fluctuation of the data. A summary of all contributions to the total systematic uncertainty is given in Table~\ref{Tab:Systematics}.

Two contributions are considered as uncertainties due to the trigger efficiency. One arises from the finite size of the simulation samples and is taken into account by varying the parameters of the efficiency function $\varepsilon_1$ within their uncertainties. The other is due to a potential discrepancy between data and simulation. This second contribution is assessed by repeating the fit using an efficiency obtained from a data sample of $\Bz \to \jpsi \KS$ decays that are topologically similar to the $b$-baryon decays of interest and reconstructed in data collected by the same trigger. In this case, extracting the efficiency from data is possible because a large sample, selected with triggers that do not bias the decay time distribution, is available (see Ref.~\cite{LHCb-PAPER-2013-065}).

\begin{table}[b!]
\caption{\small Systematic uncertainties on the lifetime measurements in fs. For the total uncertainty, all the contributions are summed in quadrature.}
\label{Tab:Systematics}
\begin{center}
\begin{tabular}{lcc}
\hline
Source                                  & $\myXb \to \jpsi \myX$ & $\myOb \to \jpsi \myO$ \\
\hline
Trigger efficiency                      & \,~9.9 & \,~6.5 \\
Reconstruction and selection efficiency &   29.0 &   45.0 \\
Signal modelling                        & \,~5.9 &   11.4 \\
Combinatorial background modelling      & \,~3.0 & \,~3.0 \\
Cross-feed background                   & \,~0.1 &   11.1 \\
Detector length scale                   & \,~0.3 & \,~0.3 \\
\hline
Total                                   &   31.4 &   48.3 \\
\hline
\end{tabular}
\end{center}
\end{table}

Two contributions are considered as uncertainties in determining the reconstruction and selection efficiency. One arises from the finite size of the simulation samples and is assessed by varying the parameter $\beta$ within its statistical uncertainty. The other takes into account the quality of the simulation of the geometrical detector acceptance, the reconstruction process and the selection. For this second contribution, the $\beta$ parameter is varied by $\pm 50\%$ to cover any possible discrepancy between data and simulation~\cite{LHCb-PAPER-2011-021}. The total systematic uncertainty related to the reconstruction and selection efficiency, taken as the quadratic sum of the two contributions, is larger for \myOb than for \myXb decays due to the larger value of the $\beta$ parameter for \myOb in 2011 data.

Several alternative fits are performed to assess the systematic uncertainties related to the signal modelling. In one fit, the Gaussian function describing the signal model in the $b$-baryon mass distribution is replaced by the sum of two Gaussian functions of common mean. The widths and the relative yields are left free. To assess the effect of the decay time resolution function, the widths of the corresponding Gaussian functions are varied by $\pm10\%$. In another alternative fit, this resolution function is taken as the sum of two Gaussian functions instead of one, where the parameters are still taken from simulation. This takes into account potential tails in the decay time resolution distribution. All variations of the function describing the decay time resolution change the fit result by a negligible amount. Therefore the systematic uncertainty related to the signal modelling is dominated by the signal description in the mass distribution.

The systematic uncertainties due to combinatorial background are taken into account with three alternative fit models. In the first, the background in the mass distribution is described with a linear function. In another fit the background is modelled with two different exponential functions for the two different years of data taking. As an alternative description of the background in the decay time distribution, three exponential functions, instead of two, are convolved with the Gaussian resolution.

The only other significant background expected is a cross-feed between the two $b$-baryon decays. The rate and mass distribution of the cross-feed backgrounds is determined by reconstructing simulated decays of one channel under the hypothesis of the other. According to simulation, 0.24 (3.0) \myOb (\myXb) decays are expected to be reconstructed as \myXb (\myOb) over the full mass range. The effect of this background on the lifetime measurement is determined by injecting simulated background events into a fit of simulated signal events and taking the observed bias as the systematic uncertainty.

The overall length scale of the vertex detector is known with a relative precision of 0.02\%~\cite{LHCb-PAPER-2013-065}. As the measured decay length is directly proportional to the overall length scale, this precision directly translates into a relative uncertainty on the lifetime measurements.

\clearpage

\section{Conclusion}

Using data samples recorded during the years 2011 and 2012, corresponding to an integrated luminosity of 3\,fb$^{-1}$, the lifetimes of the weakly decaying \myXb and \myOb baryons are measured to be

\begin{eqnarray*}
\tau (\myXb) & = & 1.55\, ^{+0.10}_{-0.09}~{\rm(stat)} \pm 0.03\,{\rm(syst)\,ps}, \\
\tau (\myOb) & = & 1.54\, ^{+0.26}_{-0.21}~{\rm(stat)} \pm 0.05\,{\rm(syst)\,ps}.
\end{eqnarray*}

These are the most precise lifetime measurements of these $b$ baryons to date. Both measurements are in agreement with the previous experimental results, in particular with the most recent ones from the \cdf collaboration of $\tau (\myXb) = 1.32 \pm 0.14$\,ps and $\tau (\myOb) = 1.66 \pm 0.47$\,ps~\cite{Aaltonen:2014al}. The measurements also lie in the range predicted by theoretical calculations~\cite{Bigi:1995, Cheng:1997, Ito:1998}.

\section*{Acknowledgements}
 
\noindent We express our gratitude to our colleagues in the CERN
accelerator departments for the excellent performance of the LHC. We
thank the technical and administrative staff at the LHCb
institutes. We acknowledge support from CERN and from the national
agencies: CAPES, CNPq, FAPERJ and FINEP (Brazil); NSFC (China);
CNRS/IN2P3 and Region Auvergne (France); BMBF, DFG, HGF and MPG
(Germany); SFI (Ireland); INFN (Italy); FOM and NWO (The Netherlands);
SCSR (Poland); MEN/IFA (Romania); MinES, Rosatom, RFBR and NRC
``Kurchatov Institute'' (Russia); MinECo, XuntaGal and GENCAT (Spain);
SNSF and SER (Switzerland); NASU (Ukraine); STFC and the Royal Society (United
Kingdom); NSF (USA). We also acknowledge the support received from EPLANET, 
Marie Curie Actions and the ERC under FP7. 
The Tier1 computing centres are supported by IN2P3 (France), KIT and BMBF (Germany),
INFN (Italy), NWO and SURF (The Netherlands), PIC (Spain), GridPP (United Kingdom).
We are indebted to the communities behind the multiple open source software packages on which we depend.
We are also thankful for the computing resources and the access to software R\&D tools provided by Yandex LLC (Russia).

\clearpage
\addcontentsline{toc}{section}{References}
\setboolean{inbibliography}{true}
\bibliographystyle{LHCb}
\bibliography{main,LHCb-PAPER,LHCb-CONF,LHCb-DP}

\ifx\mcitethebibliography\mciteundefinedmacro
\PackageError{LHCb.bst}{mciteplus.sty has not been loaded}
{This bibstyle requires the use of the mciteplus package.}\fi
\providecommand{\href}[2]{#2}
\begin{mcitethebibliography}{10}
\mciteSetBstSublistMode{n}
\mciteSetBstMaxWidthForm{subitem}{\alph{mcitesubitemcount})}
\mciteSetBstSublistLabelBeginEnd{\mcitemaxwidthsubitemform\space}
{\relax}{\relax}

\bibitem{Neubert:1998qx}
M.~Neubert, \ifthenelse{\boolean{articletitles}}{{\it {B decays and the
  Heavy-Quark Expansion}}, }{}Adv.\ Ser.\ Direct.\ High Energy Phys.\  {\bf 15}
  (1998) 239, \href{http://arxiv.org/abs/hep-ph/9702375}{{\tt
  arXiv:hep-ph/9702375}}\relax
\mciteBstWouldAddEndPuncttrue
\mciteSetBstMidEndSepPunct{\mcitedefaultmidpunct}
{\mcitedefaultendpunct}{\mcitedefaultseppunct}\relax
\EndOfBibitem
\bibitem{Buskulic:1995px}
ALEPH collaboration, D.~Buskulic {\em et~al.},
  \ifthenelse{\boolean{articletitles}}{{\it {Measurements of the $b$ baryon
  lifetime}}, }{}\href{http://dx.doi.org/10.1016/0370-2693(95)00980-Y}{Phys.\
  Lett.\  {\bf B357} (1995) 685}\relax
\mciteBstWouldAddEndPuncttrue
\mciteSetBstMidEndSepPunct{\mcitedefaultmidpunct}
{\mcitedefaultendpunct}{\mcitedefaultseppunct}\relax
\EndOfBibitem
\bibitem{Akers:1995wl}
OPAL collaboration, R.~Akers {\em et~al.},
  \ifthenelse{\boolean{articletitles}}{{\it {A measurement of the \myLb
  lifetime}}, }{}\href{http://dx.doi.org/10.1016/0370-2693(95)00553-W}{Phys.\
  Lett.\  {\bf B353} (1995) 402}\relax
\mciteBstWouldAddEndPuncttrue
\mciteSetBstMidEndSepPunct{\mcitedefaultmidpunct}
{\mcitedefaultendpunct}{\mcitedefaultseppunct}\relax
\EndOfBibitem
\bibitem{Abreu:1996py}
DELPHI collaboration, P.~Abreu {\em et~al.},
  \ifthenelse{\boolean{articletitles}}{{\it {Determination of the average
  lifetime of b baryons}}, }{}\href{http://dx.doi.org/10.1007/BF02906977}{Z.\
  Phys.\  {\bf C71} (1996) 199}\relax
\mciteBstWouldAddEndPuncttrue
\mciteSetBstMidEndSepPunct{\mcitedefaultmidpunct}
{\mcitedefaultendpunct}{\mcitedefaultseppunct}\relax
\EndOfBibitem
\bibitem{Abe:1996xl}
CDF collaboration, F.~Abe {\em et~al.},
  \ifthenelse{\boolean{articletitles}}{{\it {Measurement of \myLb lifetime
  using $\myLb \to \PLambda_c^+ \Pl^- \bar{\Pnu}$}},
  }{}\href{http://dx.doi.org/10.1103/PhysRevLett.77.1439}{Phys.\ Rev.\ Lett.\
  {\bf 77} (1996) 1439}\relax
\mciteBstWouldAddEndPuncttrue
\mciteSetBstMidEndSepPunct{\mcitedefaultmidpunct}
{\mcitedefaultendpunct}{\mcitedefaultseppunct}\relax
\EndOfBibitem
\bibitem{Altarelli:1996zb}
G.~Altarelli {\em et~al.}, \ifthenelse{\boolean{articletitles}}{{\it {Failure
  of local duality in inclusive non-leptonic heavy flavour}},
  }{}\href{http://dx.doi.org/10.1016/0370-2693(96)00637-5}{Phys.\ Lett.\  {\bf
  B382} (1996) 409}, \href{http://arxiv.org/abs/hep-ph/9604202}{{\tt
  arXiv:hep-ph/9604202}}\relax
\mciteBstWouldAddEndPuncttrue
\mciteSetBstMidEndSepPunct{\mcitedefaultmidpunct}
{\mcitedefaultendpunct}{\mcitedefaultseppunct}\relax
\EndOfBibitem
\bibitem{Cheng:1997}
H.-Y. Cheng, \ifthenelse{\boolean{articletitles}}{{\it {A phenomenological
  analysis of heavy hadron lifetimes}},
  }{}\href{http://dx.doi.org/10.1103/PhysRevD.56.2783}{Phys.\ Rev.\  {\bf D56}
  (1997) 2783}, \href{http://arxiv.org/abs/hep-ph/9704260}{{\tt
  arXiv:hep-ph/9704260}}\relax
\mciteBstWouldAddEndPuncttrue
\mciteSetBstMidEndSepPunct{\mcitedefaultmidpunct}
{\mcitedefaultendpunct}{\mcitedefaultseppunct}\relax
\EndOfBibitem
\bibitem{Ito:1998}
T.~Ito, M.~Matsuda, and Y.~Matsui, \ifthenelse{\boolean{articletitles}}{{\it
  {New possibility of solving the problem of lifetime ratio $\tau(\myLb) /
  \tau(B_d)$}}, }{}\href{http://dx.doi.org/10.1143/PTP.99.271}{Prog.\ Theor.\
  Phys.\  {\bf 99} (1998) 271}, \href{http://arxiv.org/abs/hep-ph/9705402}{{\tt
  arXiv:hep-ph/9705402}}\relax
\mciteBstWouldAddEndPuncttrue
\mciteSetBstMidEndSepPunct{\mcitedefaultmidpunct}
{\mcitedefaultendpunct}{\mcitedefaultseppunct}\relax
\EndOfBibitem
\bibitem{Uraltsev:2000vt}
N.~Uraltsev, \ifthenelse{\boolean{articletitles}}{{\it {Topics in the Heavy
  Quark Expansion}}, }{}\href{http://arxiv.org/abs/hep-ph/0010328}{{\tt
  arXiv:hep-ph/0010328}}\relax
\mciteBstWouldAddEndPuncttrue
\mciteSetBstMidEndSepPunct{\mcitedefaultmidpunct}
{\mcitedefaultendpunct}{\mcitedefaultseppunct}\relax
\EndOfBibitem
\bibitem{Gabbiani:2003xx}
F.~Gabbiani, A.~I. Onishchenko, and A.~A. Petrov,
  \ifthenelse{\boolean{articletitles}}{{\it {\myLb lifetime puzzle in
  heavy-quark expansion}},
  }{}\href{http://dx.doi.org/10.1103/PhysRevD.68.114006}{Phys.\ Rev.\  {\bf
  D68} (2003) 114006}, \href{http://arxiv.org/abs/hep-ph/0303235}{{\tt
  arXiv:hep-ph/0303235}}\relax
\mciteBstWouldAddEndPuncttrue
\mciteSetBstMidEndSepPunct{\mcitedefaultmidpunct}
{\mcitedefaultendpunct}{\mcitedefaultseppunct}\relax
\EndOfBibitem
\bibitem{Gabbiani:2004ya}
F.~Gabbiani, A.~I. Onishchenko, and A.~A. Petrov,
  \ifthenelse{\boolean{articletitles}}{{\it {Spectator effects and lifetimes of
  heavy hadrons}},
  }{}\href{http://dx.doi.org/10.1103/PhysRevD.70.094031}{Phys.\ Rev.\  {\bf
  D70} (2004) 094031}, \href{http://arxiv.org/abs/hep-ph/0407004}{{\tt
  arXiv:hep-ph/0407004}}\relax
\mciteBstWouldAddEndPuncttrue
\mciteSetBstMidEndSepPunct{\mcitedefaultmidpunct}
{\mcitedefaultendpunct}{\mcitedefaultseppunct}\relax
\EndOfBibitem
\bibitem{LHCb-PAPER-2014-003}
LHCb collaboration, R.~Aaij {\em et~al.},
  \ifthenelse{\boolean{articletitles}}{{\it {Precision measurement of the ratio
  of the $\Lambda^0_b$ to $\bar{B}^0$ lifetimes}},
  }{}\href{http://dx.doi.org/10.1016/j.physletb.2014.05.021}{Phys.\ Lett.\
  {\bf B734} (2014) 122}, \href{http://arxiv.org/abs/1402.6242}{{\tt
  arXiv:1402.6242}}\relax
\mciteBstWouldAddEndPuncttrue
\mciteSetBstMidEndSepPunct{\mcitedefaultmidpunct}
{\mcitedefaultendpunct}{\mcitedefaultseppunct}\relax
\EndOfBibitem
\bibitem{Bigi:1995}
I.~I. Bigi, \ifthenelse{\boolean{articletitles}}{{\it {The QCD perspective on
  lifetimes of heavy-flavour hadrons}},
  }{}\href{http://arxiv.org/abs/hep-ph/9508408}{{\tt
  arXiv:hep-ph/9508408}}\relax
\mciteBstWouldAddEndPuncttrue
\mciteSetBstMidEndSepPunct{\mcitedefaultmidpunct}
{\mcitedefaultendpunct}{\mcitedefaultseppunct}\relax
\EndOfBibitem
\bibitem{Aaltonen:2007ap}
CDF collaboration, T.~Aaltonen {\em et~al.},
  \ifthenelse{\boolean{articletitles}}{{\it {Observation and mass measurement
  of the baryon \myXb}},
  }{}\href{http://dx.doi.org/10.1103/PhysRevLett.99.052002}{Phys.\ Rev.\ Lett.\
   {\bf 99} (2007) 052002}, \href{http://arxiv.org/abs/0707.0589}{{\tt
  arXiv:0707.0589}}\relax
\mciteBstWouldAddEndPuncttrue
\mciteSetBstMidEndSepPunct{\mcitedefaultmidpunct}
{\mcitedefaultendpunct}{\mcitedefaultseppunct}\relax
\EndOfBibitem
\bibitem{Aaltonen:2009ny}
CDF collaboration, T.~Aaltonen {\em et~al.},
  \ifthenelse{\boolean{articletitles}}{{\it {Observation of the \myOb baryon
  and measurement of the properties of the \myXb and \myOb baryons}},
  }{}\href{http://dx.doi.org/10.1103/PhysRevD.80.072003}{Phys.\ Rev.\  {\bf
  D80} (2009) 072003}, \href{http://arxiv.org/abs/0905.3123}{{\tt
  arXiv:0905.3123}}\relax
\mciteBstWouldAddEndPuncttrue
\mciteSetBstMidEndSepPunct{\mcitedefaultmidpunct}
{\mcitedefaultendpunct}{\mcitedefaultseppunct}\relax
\EndOfBibitem
\bibitem{Abazov:2007am}
D0 collaboration, V.~Abazov {\em et~al.},
  \ifthenelse{\boolean{articletitles}}{{\it {Direct observation of the strange
  $b$ baryon \myXb}},
  }{}\href{http://dx.doi.org/10.1103/PhysRevLett.99.052001}{Phys.\ Rev.\ Lett.\
   {\bf 99} (2007) 052001}, \href{http://arxiv.org/abs/0706.1690}{{\tt
  arXiv:0706.1690}}\relax
\mciteBstWouldAddEndPuncttrue
\mciteSetBstMidEndSepPunct{\mcitedefaultmidpunct}
{\mcitedefaultendpunct}{\mcitedefaultseppunct}\relax
\EndOfBibitem
\bibitem{Abazov:2008qm}
D0 collaboration, V.~Abazov {\em et~al.},
  \ifthenelse{\boolean{articletitles}}{{\it {Observation of the doubly strange
  $b$ baryon \myOb}},
  }{}\href{http://dx.doi.org/10.1103/PhysRevLett.101.232002}{Phys.\ Rev.\
  Lett.\  {\bf 101} (2008) 232002}, \href{http://arxiv.org/abs/0808.4142}{{\tt
  arXiv:0808.4142}}\relax
\mciteBstWouldAddEndPuncttrue
\mciteSetBstMidEndSepPunct{\mcitedefaultmidpunct}
{\mcitedefaultendpunct}{\mcitedefaultseppunct}\relax
\EndOfBibitem
\bibitem{Abreu:1995qq}
DELPHI collaboration, P.~Abreu {\em et~al.},
  \ifthenelse{\boolean{articletitles}}{{\it {Production of strange B baryons
  decaying into $\PXi^{\mp}$ - $\Pl^{\mp}$ pairs at LEP}},
  }{}\href{http://dx.doi.org/10.1007/BF01565255}{Z.\ Phys.\  {\bf C68} (1995)
  541}\relax
\mciteBstWouldAddEndPuncttrue
\mciteSetBstMidEndSepPunct{\mcitedefaultmidpunct}
{\mcitedefaultendpunct}{\mcitedefaultseppunct}\relax
\EndOfBibitem
\bibitem{Abdallah:2005xi}
DELPHI collaboration, J.~Abdallah {\em et~al.},
  \ifthenelse{\boolean{articletitles}}{{\it {Production of $\PXi^{0}_{c}$ and
  $\PXi_{b}$ in $Z$ decays and lifetime measurement of $\PXi_{b}$}},
  }{}\href{http://dx.doi.org/10.1140/epjc/s2005-02388-4}{Eur.\ Phys.\ J.\  {\bf
  C44} (2005) 299}, \href{http://arxiv.org/abs/hep-ex/0510023}{{\tt
  arXiv:hep-ex/0510023}}\relax
\mciteBstWouldAddEndPuncttrue
\mciteSetBstMidEndSepPunct{\mcitedefaultmidpunct}
{\mcitedefaultendpunct}{\mcitedefaultseppunct}\relax
\EndOfBibitem
\bibitem{Buskulic:1996zy}
ALEPH collaboration, D.~Buskulic {\em et~al.},
  \ifthenelse{\boolean{articletitles}}{{\it {Strange $b$ baryon production and
  lifetime in $Z$ decays}},
  }{}\href{http://dx.doi.org/10.1016/0370-2693(96)00925-2}{Phys.\ Lett.\  {\bf
  B384} (1996) 449}\relax
\mciteBstWouldAddEndPuncttrue
\mciteSetBstMidEndSepPunct{\mcitedefaultmidpunct}
{\mcitedefaultendpunct}{\mcitedefaultseppunct}\relax
\EndOfBibitem
\bibitem{Aaltonen:2014al}
CDF collaboration, T.~Aaltonen {\em et~al.},
  \ifthenelse{\boolean{articletitles}}{{\it {Mass and lifetime measurements of
  bottom and charm baryons in $p\bar{p}$ collisions at $\sqrt{s} = 1.96$~TeV}},
  }{}\href{http://dx.doi.org/10.1103/PhysRevD.89.072014}{Phys.\ Rev.\  {\bf
  D89} (2014) 072014}, \href{http://arxiv.org/abs/1403.8126}{{\tt
  arXiv:1403.8126}}\relax
\mciteBstWouldAddEndPuncttrue
\mciteSetBstMidEndSepPunct{\mcitedefaultmidpunct}
{\mcitedefaultendpunct}{\mcitedefaultseppunct}\relax
\EndOfBibitem
\bibitem{LHCb-PAPER-2012-048}
LHCb collaboration, R.~Aaij {\em et~al.},
  \ifthenelse{\boolean{articletitles}}{{\it {Measurements of the $\Lambda_b^0$,
  $\Xi_b^-$ and $\Omega_b^-$ baryon masses}},
  }{}\href{http://dx.doi.org/10.1103/PhysRevLett.110.182001}{Phys.\ Rev.\
  Lett.\  {\bf 110} (2013) 182001}, \href{http://arxiv.org/abs/1302.1072}{{\tt
  arXiv:1302.1072}}\relax
\mciteBstWouldAddEndPuncttrue
\mciteSetBstMidEndSepPunct{\mcitedefaultmidpunct}
{\mcitedefaultendpunct}{\mcitedefaultseppunct}\relax
\EndOfBibitem
\bibitem{LHCb-PAPER-2013-065}
LHCb collaboration, R.~Aaij {\em et~al.},
  \ifthenelse{\boolean{articletitles}}{{\it {Measurements of the $B^+$, $B^0$,
  $B_s^0$ meson and $\Lambda_b^0$ baryon lifetimes}},
  }{}\href{http://dx.doi.org/10.1007/JHEP04(2014)114}{JHEP {\bf 04} (2014)
  114}, \href{http://arxiv.org/abs/1402.2554}{{\tt arXiv:1402.2554}}\relax
\mciteBstWouldAddEndPuncttrue
\mciteSetBstMidEndSepPunct{\mcitedefaultmidpunct}
{\mcitedefaultendpunct}{\mcitedefaultseppunct}\relax
\EndOfBibitem
\bibitem{Alves:2008zz}
LHCb collaboration, A.~A. Alves~Jr. {\em et~al.},
  \ifthenelse{\boolean{articletitles}}{{\it {The \lhcb detector at the LHC}},
  }{}\href{http://dx.doi.org/10.1088/1748-0221/3/08/S08005}{JINST {\bf 3}
  (2008) S08005}\relax
\mciteBstWouldAddEndPuncttrue
\mciteSetBstMidEndSepPunct{\mcitedefaultmidpunct}
{\mcitedefaultendpunct}{\mcitedefaultseppunct}\relax
\EndOfBibitem
\bibitem{LHCb-DP-2013-003}
R.~Arink {\em et~al.}, \ifthenelse{\boolean{articletitles}}{{\it {Performance
  of the LHCb Outer Tracker}},
  }{}\href{http://dx.doi.org/10.1088/1748-0221/9/01/P01002}{JINST {\bf 9}
  (2014) P01002}, \href{http://arxiv.org/abs/1311.3893}{{\tt
  arXiv:1311.3893}}\relax
\mciteBstWouldAddEndPuncttrue
\mciteSetBstMidEndSepPunct{\mcitedefaultmidpunct}
{\mcitedefaultendpunct}{\mcitedefaultseppunct}\relax
\EndOfBibitem
\bibitem{LHCb-DP-2012-003}
M.~Adinolfi {\em et~al.}, \ifthenelse{\boolean{articletitles}}{{\it
  {Performance of the \lhcb RICH detector at the LHC}},
  }{}\href{http://dx.doi.org/10.1140/epjc/s10052-013-2431-9}{Eur.\ Phys.\ J.\
  {\bf C73} (2013) 2431}, \href{http://arxiv.org/abs/1211.6759}{{\tt
  arXiv:1211.6759}}\relax
\mciteBstWouldAddEndPuncttrue
\mciteSetBstMidEndSepPunct{\mcitedefaultmidpunct}
{\mcitedefaultendpunct}{\mcitedefaultseppunct}\relax
\EndOfBibitem
\bibitem{LHCb-DP-2012-002}
A.~A. Alves~Jr. {\em et~al.}, \ifthenelse{\boolean{articletitles}}{{\it
  {Performance of the LHCb muon system}},
  }{}\href{http://dx.doi.org/10.1088/1748-0221/8/02/P02022}{JINST {\bf 8}
  (2013) P02022}, \href{http://arxiv.org/abs/1211.1346}{{\tt
  arXiv:1211.1346}}\relax
\mciteBstWouldAddEndPuncttrue
\mciteSetBstMidEndSepPunct{\mcitedefaultmidpunct}
{\mcitedefaultendpunct}{\mcitedefaultseppunct}\relax
\EndOfBibitem
\bibitem{Sjostrand:2006za}
T.~Sj\"{o}strand, S.~Mrenna, and P.~Skands,
  \ifthenelse{\boolean{articletitles}}{{\it {PYTHIA 6.4 physics and manual}},
  }{}\href{http://dx.doi.org/10.1088/1126-6708/2006/05/026}{JHEP {\bf 05}
  (2006) 026}, \href{http://arxiv.org/abs/hep-ph/0603175}{{\tt
  arXiv:hep-ph/0603175}}\relax
\mciteBstWouldAddEndPuncttrue
\mciteSetBstMidEndSepPunct{\mcitedefaultmidpunct}
{\mcitedefaultendpunct}{\mcitedefaultseppunct}\relax
\EndOfBibitem
\bibitem{Sjostrand:2007gs}
T.~Sj\"{o}strand, S.~Mrenna, and P.~Skands,
  \ifthenelse{\boolean{articletitles}}{{\it {A brief introduction to PYTHIA
  8.1}}, }{}\href{http://dx.doi.org/10.1016/j.cpc.2008.01.036}{Comput.\ Phys.\
  Commun.\  {\bf 178} (2008) 852}, \href{http://arxiv.org/abs/0710.3820}{{\tt
  arXiv:0710.3820}}\relax
\mciteBstWouldAddEndPuncttrue
\mciteSetBstMidEndSepPunct{\mcitedefaultmidpunct}
{\mcitedefaultendpunct}{\mcitedefaultseppunct}\relax
\EndOfBibitem
\bibitem{LHCb-PROC-2010-056}
I.~Belyaev {\em et~al.}, \ifthenelse{\boolean{articletitles}}{{\it {Handling of
  the generation of primary events in \gauss, the \lhcb simulation framework}},
  }{}\href{http://dx.doi.org/10.1109/NSSMIC.2010.5873949}{Nuclear Science
  Symposium Conference Record (NSS/MIC) {\bf IEEE} (2010) 1155}\relax
\mciteBstWouldAddEndPuncttrue
\mciteSetBstMidEndSepPunct{\mcitedefaultmidpunct}
{\mcitedefaultendpunct}{\mcitedefaultseppunct}\relax
\EndOfBibitem
\bibitem{Lange:2001uf}
D.~J. Lange, \ifthenelse{\boolean{articletitles}}{{\it {The EvtGen particle
  decay simulation package}},
  }{}\href{http://dx.doi.org/10.1016/S0168-9002(01)00089-4}{Nucl.\ Instrum.\
  Meth.\  {\bf A462} (2001) 152}\relax
\mciteBstWouldAddEndPuncttrue
\mciteSetBstMidEndSepPunct{\mcitedefaultmidpunct}
{\mcitedefaultendpunct}{\mcitedefaultseppunct}\relax
\EndOfBibitem
\bibitem{Golonka:2005pn}
P.~Golonka and Z.~Was, \ifthenelse{\boolean{articletitles}}{{\it {PHOTOS Monte
  Carlo: a precision tool for QED corrections in $Z$ and $W$ decays}},
  }{}\href{http://dx.doi.org/10.1140/epjc/s2005-02396-4}{Eur.\ Phys.\ J.\  {\bf
  C45} (2006) 97}, \href{http://arxiv.org/abs/hep-ph/0506026}{{\tt
  arXiv:hep-ph/0506026}}\relax
\mciteBstWouldAddEndPuncttrue
\mciteSetBstMidEndSepPunct{\mcitedefaultmidpunct}
{\mcitedefaultendpunct}{\mcitedefaultseppunct}\relax
\EndOfBibitem
\bibitem{Allison:2006ve}
Geant4 collaboration, J.~Allison {\em et~al.},
  \ifthenelse{\boolean{articletitles}}{{\it {Geant4 developments and
  applications}}, }{}\href{http://dx.doi.org/10.1109/TNS.2006.869826}{IEEE
  Trans.\ Nucl.\ Sci.\  {\bf 53} (2006) 270}\relax
\mciteBstWouldAddEndPuncttrue
\mciteSetBstMidEndSepPunct{\mcitedefaultmidpunct}
{\mcitedefaultendpunct}{\mcitedefaultseppunct}\relax
\EndOfBibitem
\bibitem{Agostinelli:2002hh}
Geant4 collaboration, S.~Agostinelli {\em et~al.},
  \ifthenelse{\boolean{articletitles}}{{\it {Geant4: a simulation toolkit}},
  }{}\href{http://dx.doi.org/10.1016/S0168-9002(03)01368-8}{Nucl.\ Instrum.\
  Meth.\  {\bf A506} (2003) 250}\relax
\mciteBstWouldAddEndPuncttrue
\mciteSetBstMidEndSepPunct{\mcitedefaultmidpunct}
{\mcitedefaultendpunct}{\mcitedefaultseppunct}\relax
\EndOfBibitem
\bibitem{LHCb-PROC-2011-006}
M.~Clemencic {\em et~al.}, \ifthenelse{\boolean{articletitles}}{{\it {The \lhcb
  simulation application, \gauss: design, evolution and experience}},
  }{}\href{http://dx.doi.org/10.1088/1742-6596/331/3/032023}{{J.\ Phys.\ Conf.\
  Ser.\ } {\bf 331} (2011) 032023}\relax
\mciteBstWouldAddEndPuncttrue
\mciteSetBstMidEndSepPunct{\mcitedefaultmidpunct}
{\mcitedefaultendpunct}{\mcitedefaultseppunct}\relax
\EndOfBibitem
\bibitem{PDG2012}
Particle Data Group, J.~Beringer {\em et~al.},
  \ifthenelse{\boolean{articletitles}}{{\it {\href{http://pdg.lbl.gov/}{Review
  of particle physics}}},
  }{}\href{http://dx.doi.org/10.1103/PhysRevD.86.010001}{Phys.\ Rev.\  {\bf
  D86} (2012) 010001}, {and 2013 partial update for the 2014 edition}\relax
\mciteBstWouldAddEndPuncttrue
\mciteSetBstMidEndSepPunct{\mcitedefaultmidpunct}
{\mcitedefaultendpunct}{\mcitedefaultseppunct}\relax
\EndOfBibitem
\bibitem{Hulsbergen:2005pu}
W.~D. Hulsbergen, \ifthenelse{\boolean{articletitles}}{{\it {Decay chain
  fitting with a Kalman filter}},
  }{}\href{http://dx.doi.org/10.1016/j.nima.2005.06.078}{Nucl.\ Instrum.\
  Meth.\  {\bf A552} (2005) 566},
  \href{http://arxiv.org/abs/physics/0503191}{{\tt
  arXiv:physics/0503191}}\relax
\mciteBstWouldAddEndPuncttrue
\mciteSetBstMidEndSepPunct{\mcitedefaultmidpunct}
{\mcitedefaultendpunct}{\mcitedefaultseppunct}\relax
\EndOfBibitem
\bibitem{LHCb-PAPER-2011-021}
LHCb collaboration, R.~Aaij {\em et~al.},
  \ifthenelse{\boolean{articletitles}}{{\it {Measurement of the $CP$-violating
  phase $\phi_s$ in the decay $B^0_s \to J/\psi \phi$}},
  }{}\href{http://dx.doi.org/10.1103/PhysRevLett.108.101803}{Phys.\ Rev.\
  Lett.\  {\bf 108} (2012) 101803}, \href{http://arxiv.org/abs/1112.3183}{{\tt
  arXiv:1112.3183}}\relax
\mciteBstWouldAddEndPuncttrue
\mciteSetBstMidEndSepPunct{\mcitedefaultmidpunct}
{\mcitedefaultendpunct}{\mcitedefaultseppunct}\relax
\EndOfBibitem
\end{mcitethebibliography}

\end{document}